\documentclass[twocolumn]{aastex631}

\usepackage{xspace}
\usepackage{CJK}
\usepackage{rotating}
\usepackage{afterpage}
\usepackage{changepage}
\usepackage{sidecap}
\usepackage{comment}

\hypersetup{
    colorlinks=true,
    linkcolor=blue,
    filecolor=magenta,      
    urlcolor=cyan,
}

\usepackage[normalem]{ulem} 
\usepackage{cancel} 
\usepackage{soul}

\usepackage{color}

\newcommand{\kms}{\ensuremath{\rm{km\,s}^{-1}}\xspace}
\newcommand{\alphaco}{\ensuremath{\alpha_{\rm{CO}}}\xspace}

\newcommand{\uJybeam}{\ensuremath{\mu\rm{Jy}\,\rm{beam}^{-1}}\xspace}

\newcommand{\Mstar}{\ensuremath{M_{\rm{*}}}\xspace}

\newcommand{\Msol}{\ensuremath{\rm{M}_\odot}\xspace}
\newcommand{\sfr}{\ensuremath{\rm{M}_\odot\,yr^{-1}}\xspace}

\newcommand{\etal}{et~al.\xspace}

\newcommand{\arc}{\ensuremath{''}\xspace}
\newcommand{\squiggle}{SQuIGG\ensuremath{\vec{L}}E\xspace}
\newcommand{\PSBI}{SDSS\,J1448+1010\xspace}
\newcommand{\PSBII}{SDSS\,J2258+2313\xspace}

\shortauthors{D'Onofrio, et~al.}
\shorttitle{Quenching Through Tidal Gas Removal}

\begin{document}
\begin{CJK*}{UTF8}{gbsn}
\defcitealias{spilker_star_2022}{S22}

\title{Quenching Through Tidal Gas Removal:\\Molecular Gas and Star Formation in Tidal Tails of $z\sim0.7$ Post-Starburst Galaxies}

\correspondingauthor{Vincenzo R. D'Onofrio}
\email{donofr19@tamu.edu}

\author[0000-0002-1759-6205]{Vincenzo~R.~D'Onofrio}
\affiliation{Department of Physics and Astronomy and George P. and Cynthia Woods Mitchell Institute for Fundamental Physics and Astronomy, Texas A\&M University, 4242 TAMU, College Station, TX 77843-4242, US}

\author[0000-0003-3256-5615]{Justin~S.~Spilker}
\affiliation{Department of Physics and Astronomy and George P. and Cynthia Woods Mitchell Institute for Fundamental Physics and Astronomy, Texas A\&M University, 4242 TAMU, College Station, TX 77843-4242, US}

\author[0000-0001-5063-8254]{Rachel~Bezanson}
\affiliation{Department of Physics and Astronomy and PITT PACC, University of Pittsburgh, Pittsburgh, PA 15260, USA}

\author[0000-0002-1109-1919]{Robert~Feldmann}
\affiliation{Department of Astrophysics, University of Zurich, Winterthurerstrasse 190, Zurich CH-8057, Switzerland}

\author[0000-0003-4700-663X]{Andy~D.~Goulding}
\affiliation{Department of Astrophysical Sciences, Princeton University, Princeton, NJ 08544, USA}

\author[0000-0002-5612-3427]{Jenny~E.~Greene}
\affiliation{Department of Astrophysical Sciences, Princeton University, Princeton, NJ 08544, USA}

\author[0000-0002-7613-9872]{Mariska~Kriek}
\affiliation{Leiden Observatory, Leiden University, P.O. Box 9513, 2300 RA Leiden, The Netherlands}

\author[0000-0002-0696-6952]{Yuanze~Luo}
\affiliation{Department of Physics and Astronomy and George P. and Cynthia Woods Mitchell Institute for Fundamental Physics and Astronomy, Texas A\&M University, 4242 TAMU, College Station, TX 77843-4242, US}

\author[0000-0002-7064-4309]{Desika~Narayanan}
\affiliation{Department of Astronomy, University of Florida, 211 Bryant Space Science Center, Gainesville, FL 32611, USA}

\author[0000-0003-4075-7393]{David~J.~Setton}\thanks{Brinson Prize Fellow}
\affiliation{Department of Astrophysical Sciences, Princeton University, Princeton, NJ 08544, USA}

\author[0000-0002-1714-1905]{Katherine~A.~Suess}
\affiliation{Department for Astrophysical \& Planetary Science, University of Colorado, Boulder, CO 80309, USA}

\author[0000-0001-6454-1699]{Yunchong~Zhang}
\affiliation{Department of Physics and Astronomy and PITT PACC, University of Pittsburgh, Pittsburgh, PA 15260, USA}

\author[0000-0002-6768-8335]{Pengpei~Zhu}
\affiliation{Cosmic Dawn Center (DAWN), Denmark}
\affiliation{DTU Space, Technical University of Denmark, Elektrovej 327, 2800 Kgs. Lyngby, Denmark}


\begin{abstract}

\noindent 

The active suppression of star formation in galaxies is critical in preventing the growth of overly massive systems and explaining the formation of present-day elliptical galaxies. We present a high-resolution, spatially-resolved multiwavelength study of two $z\sim0.7$ massive post-starburst galaxies, \PSBI and \PSBII, from the \squiggle survey (Studying Quenching in Intermediate-z Galaxies: Gas, angu$\vec{L}$ar momentum, and Evolution), providing new insights into the role of mergers in driving quenching. ALMA CO(2--1) observations show that both galaxies removed $\sim$50\% of their molecular gas into extended tidal tails, spanning up to 65\,kpc, following recent mergers. HST WFC3 imaging and grism spectroscopy show that while \PSBI exhibits H$\alpha$ emission in its northern tidal tail consistent with ongoing star formation, \PSBII lacks detectable star-forming activity outside the central galaxy. VLA 6\,GHz continuum data reveal compact radio emission in \PSBII, while \PSBI hosts small radio jets indicative of AGN activity. Both galaxies retain substantial molecular gas reservoirs in their central regions that appear more turbulent than `normal' star-forming galaxies, likely contributing to the observed low star formation rates in the hosts. Despite similarities in their cold gas content and tidal features the galaxies are distinct from each other in their star formation, gas--star alignment, and radio morphology, highlighting the complexity of tidal gas removal as a quenching mechanism at intermediate redshifts.      

\end{abstract}

\section{Introduction} \label{intro}
Present-day galaxies have long been known to come in two main flavors: blue disky galaxies with ongoing star formation and red elliptical galaxies with old stellar populations \citep[e.g.,][]{kauffmann_dependence_2003}. However, we still do not fully understand the physical processes responsible for this dichotomy. In order to produce galaxy populations that match with even the most fundamental observational characteristics, modern galaxy formation simulations require the introduction of additional energetic `feedback' processes. Several physical mechanisms have been proposed to attempt to explain the rapid suppression of star formation (or `quenching') via processes that are capable of removing, exhausting, or heating the cold molecular gas reserves in which stars form. Feedback from active galactic nuclei (AGN) is commonly invoked as a process capable of quenching star formation by removing cold gas from the galaxy \citep[e.g.,][]{di_matteo_energy_2005, croton_many_2006, hopkins_unified_2006}, potentially induced by the effects of a major merger \citep[e.g.,][]{springel_modelling_2005, wellons_formation_2015}. Likewise, \citet{darvish_effects_2016} suggest that dense local environments can more efficiently quench galaxies, perhaps due to a higher merger rate of massive systems. Massive galaxy-scale molecular outflows, driven not only by AGN but also supernovae, have been observed across cosmic epochs with the energetics required to play a role in the rapid suppression of star formation \citep[e.g.,][]{alatalo_escape_2015, fluetsch_cold_2019, spilker_ubiquitous_2020}. Alternatively, the accretion of cold gas onto the galaxy could decline as a result of virial shocks \citep[e.g.,][]{keres_how_2005, dekel_galaxy_2006} or with the increasing mass of the dark matter halo, which can suppress further gas inflow \citep[e.g.,][]{croton_many_2006, feldmann_argo_2015, feldmann_colours_2017}. Although these mechanisms are undoubtedly important to our picture of the galaxy formation process, there may also be additional physical processes that aid quenching across cosmic time.

These various quenching mechanisms leave different signatures on the stars and gas in galaxies. The molecular gas properties, in particular, may be crucial to understand which physical mechanisms are responsible for quenching, because molecular gas is the direct fuel for star formation. Post-starburst galaxies (PSBs) serve as ideal laboratories for this study. The A-star-dominated optical spectra of PSBs indicate an intense period of star formation followed by a rapid quenching event $\sim$0.1-1\,Gyr ago, providing an uncontaminated view of quenching (see \citet{french_evolution_2021} for a recent review of PSBs). While rare at all redshifts, PSBs contribute to $\sim$5\% of the total galaxy population by $z \sim 2$, nearly $50\times$ more common than those locally \citep{wild_evolution_2016}. At $z \sim 2$, these fast quenching systems account for about half of the growth of the red sequence \citep{belli_mosfire_2019}. Furthermore, the burst phase of local PSBs typically adds only $\sim$5\% to the pre-existing stellar mass \citep[e.g.,][]{zabludoff_environment_1996}, while starbursts at higher redshifts can form upwards of $\sim$70\% \citep[e.g.,][]{suess_recovering_2022} before rapidly quenching, thus representing the primary formation epoch of these galaxies.

The \squiggle program (Studying Quenching in Intermediate-z Galaxies: Gas, angu$\vec{L}$ar momentum, and Evolution) \citep{suess_mathrmsquiggecle_2022} selected 1318 recently-quenched, massive PSBs at $z\sim0.7$ from the Sloan Digital Sky Survey (SDSS) Data Release 14 (DR14) spectroscopic database \citep{abolfathi_fourteenth_2018}. \squiggle is designed to identify PSBs that recently quenched their primary star formation epoch, targeting galaxies distant enough to represent the high-redshift post-starburst population but nearby enough to allow for detailed analysis. A detailed description of the spectroscopic identification and stellar populations of the \squiggle PSBs can be found in \citet{suess_mathrmsquiggecle_2022}. From a representative sample of \squiggle galaxies with flat radial age gradients, we infer that quenching likely occurs uniformly throughout these galaxies \citep{setton_squigg_2020}. Visual classifications of merger-induced tidal disturbances of \squiggle galaxies indicate that merger features such as tidal disturbances are associated with younger PSBs \citep{verrico_merger_2023}. Younger \squiggle PSBs are $\sim$10 times more likely to host optical nuclear activity, while radio activity shows only a weak association with stellar mass and no clear dependence on stellar age \citep{greene_role_2020}. Prior Atacama Large Millimeter Array (ALMA) CO(2--1) observations of \squiggle PSBs suggest that the quenching of star formation precedes the removal of cold gas reservoirs. However, the gas reservoirs disappear rapidly within the following $100-200$\,Myr post-quenching, requiring additional heating or removal, suggesting that quenching does not require the total removal of cold gas \citep{suess_massive_2017, bezanson_now_2022}. \citet{spilker_star_2022} (hereafter S22) previously presented the discovery of vast molecular gas features in the tidal tail regions of a massive $z=0.646$ \squiggle post-starburst SDSS\,J144845.91+101010.5 (\PSBI), extending far beyond the central galaxy. The tidal removal of the cold molecular gas was suggested to be the proximate cause of quenching in \PSBI. 

\begin{figure*}[ht!]
\centering{
\includegraphics[width=0.9\textwidth]{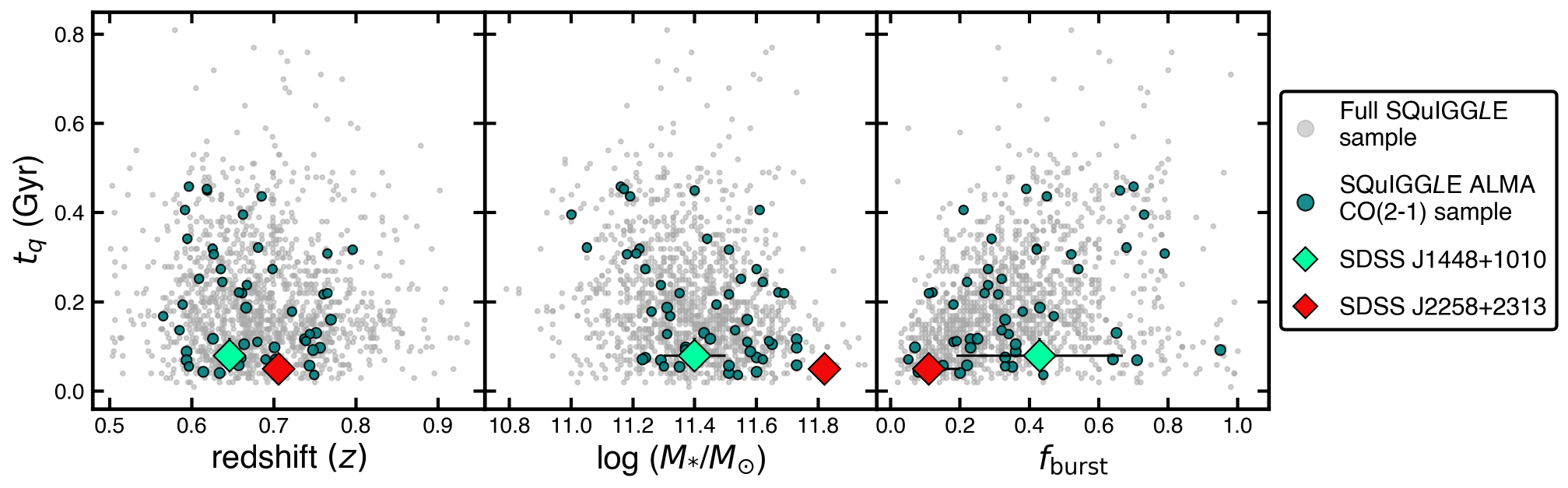}
}
\caption{The distribution of time since quenching against redshift (left), stellar mass (middle), and the fraction of the mass formed in the recent burst (right). In each panel, the full \squiggle sample \citep{suess_mathrmsquiggecle_2022} is indicated by the gray circles with the \squiggle ALMA CO(2--1) subsample (\citealt{bezanson_now_2022}; Setton et al. in prep.) shown in green circles. The targets presented in this work \PSBI and \PSBII are shown in light green and red diamonds, respectively. These post-starbursts are younger and more massive than most objects in the \squiggle sample.  
}\label{fig:compToSquiggle}
\end{figure*}

In this work, we present a second such example of vast molecular tidal gas characteristics in SDSS\,J225805.67+231316.1 (\PSBII), a massive $z=0.706$ post-starburst from the \squiggle sample. We also perform a multiwavelength analysis of both galaxies to understand the physical mechanisms responsible for the suppression of star formation. In Section~\ref{data}, we describe the target galaxies and observational data, including ALMA data and follow-up Hubble Space Telescope (HST) Wide Field Camera 3 (WFC3) grism and Karl G. Jansky Very Large Array (VLA) observations. Section~\ref{analysis} describes our analysis of these datasets, including possible limitations due to the assumed CO-H$_2$ conversion factor $\alphaco$ and the identification of AGN in each galaxy. In Section~\ref{resultstails}, we present the results of the extended molecular gas features, assessing their tidal origins and connection to the recent merger history. In Section~\ref{resultscentrals}, we explore the suppression of star formation in the central galaxies, investigating the role of turbulence and AGN feedback in maintaining their quenched state. Finally, we conclude our work and discuss future directions in Section~\ref{conclusions}. Throughout this paper we assume a \citet{chabrier_galactic_2003} initial mass function and a concordance $\Lambda$ cold dark matter ($\Lambda$CDM) cosmology with $\Omega_m=0.3$ and $h=0.7$.  

\section{Data and Methods} \label{data}

\subsection{Target Post-Starburst Galaxies} \label{targets}
The two PSBs in this study were selected from a larger ALMA CO(2--1) survey of 51 galaxies from the \squiggle sample. The full sample will be presented in Setton \etal (in prep.). Table \ref{tab:sedProps} summarizes basic properties of both galaxies derived from spectrophotometric model fitting \citep[\citetalias{spilker_star_2022};][]{suess_mathrmsquiggecle_2022}. We define $t_q$ as the time since the onset of quenching and $f_\mathrm{burst}$ as the fraction of the total stellar mass formed in a recent burst of star formation; see \citet{suess_recovering_2022} for more detail on these definitions. Figure \ref{fig:compToSquiggle} compares our two targets with both the full \squiggle sample and the subset with ALMA follow-up. Compared to the rest of the PSBs from \squiggle both galaxies are younger than the typical galaxy in the parent sample. Despite being skewed towards the massive end of the full sample, the amount of mass formed from the recent starburst is relatively small. However, even a $f_\mathrm{burst}\sim10\%$ is still twice the fraction typically found in local PSBs \citep[e.g.,][]{zabludoff_environment_1996}. 

\begin{deluxetable}{lcc}[b!]
    \tablecaption{Results from SED-fitting for \squiggle PSBs \PSBI and \PSBII. \label{tab:sedProps}}
    \tablecolumns{3}
    \tablewidth{0pt}
    \tablehead{
        \colhead{} & \colhead{J1448$+$1010\tablenotemark{a}} & \colhead{J2258$+$2313\tablenotemark{b}}
    }
    \startdata
    $z$ & $0.646$ & $0.706$ \\
    $\log \Mstar/\Msol$ & $11.40_{-0.10}^{+0.10}$ & $11.82_{-0.05}^{+0.03}$ \\
    $t_q$ (Gyr) & $0.08_{-0.03}^{+0.04}$ & $0.05_{-0.03}^{+0.03}$ \\
    $f_\mathrm{burst}$ & $0.43_{-0.24}^{+0.21}$ & $0.11_{-0.03}^{+0.09}$ \\
    \enddata
    \tablenotetext{a}{Results are from \citetalias{spilker_star_2022} which include SDSS photometry and spectroscopy and WISE, WFC3/F110W, and FLAMINGOS-2 $H$-band photometry.}
    \tablenotetext{b}{Results are from \citet{suess_mathrmsquiggecle_2022} which include SDSS photometry and spectroscopy and WISE photometry.}
\end{deluxetable}

\subsection{ALMA Observations} \label{almaobs}
To investigate the molecular gas properties and spatial distribution of cold gas in these PSBs, we obtained ALMA CO(2--1) observations (\citealt{bezanson_now_2022}; Setton \etal in prep.). As discussed in \citetalias{spilker_star_2022}, follow-up deep low-resolution observations of \PSBI (PI: Suess, Program \#2018.1.01264.S) were carried out to better understand the distribution of CO emission, particularly the low-surface-brightness extragalactic emission. The initial CO(2--1) observations of \PSBII were delivered at low-resolution ${\sim}1.2\arc$, allowing for the detection of the large-scale faint emission. We acquired higher-resolution ${\sim}0.5\arc$ observations to measure the spatial extent of the CO(2--1) emission in this object, with total on source time of 80\,minutes (PI: Suess, Program \#2019.1.00221.S).  

\begin{figure*}[ht!]
\centering{
\includegraphics[width=0.99\textwidth]{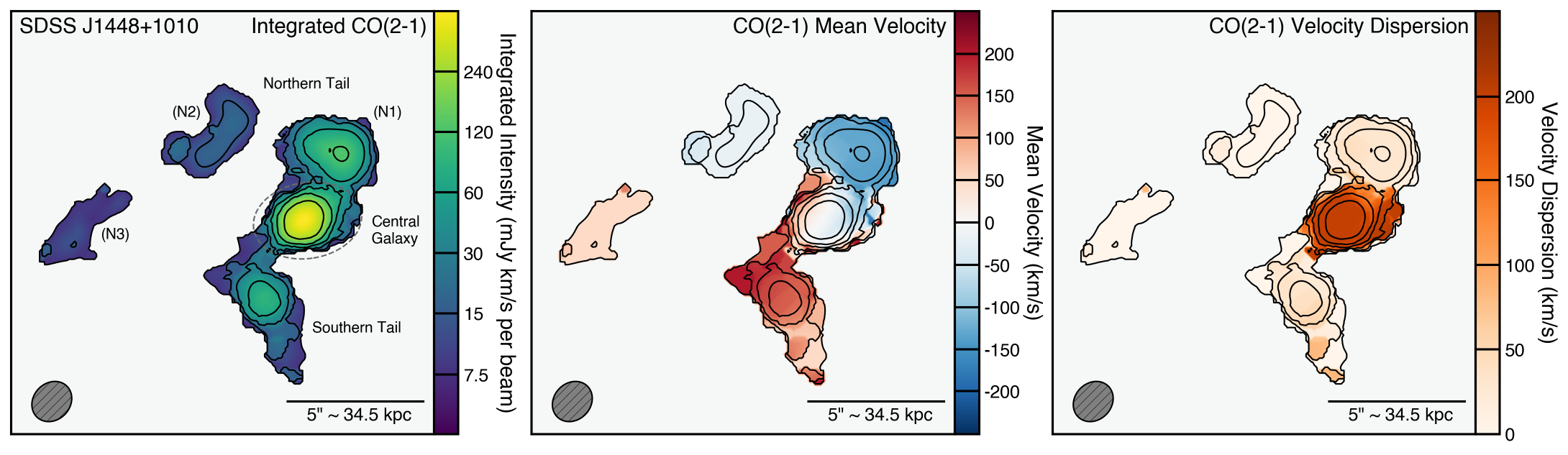}
\includegraphics[width=0.85\textwidth]{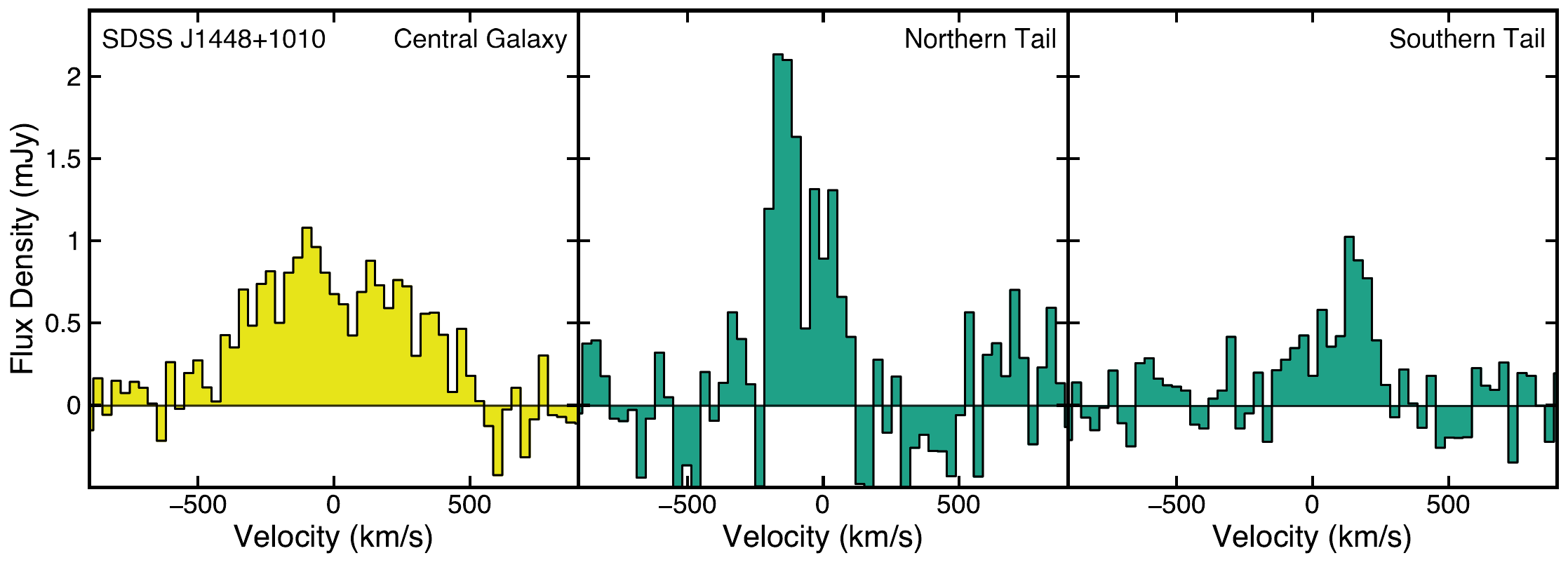}
}
\caption{Moment maps and spectra of ALMA CO(2--1) emission for \PSBI, reproduced from \citetalias{spilker_star_2022}. Top row: Integrated CO(2--1) emission (left), mean velocity (center), and velocity dispersion (right). Contours begin at $3\sigma$ and increase in powers of two, with the central galaxy region highlighted by the dashed ellipse. CO emission extends up to $\sim$65 kpc, with blueshifted northern and redshifted southern tidal features offset by $\approx100-150$\,\kms from the systemic velocity. The central galaxy exhibits turbulent gas with velocity dispersions up to $\sim$250\,\kms. Bottom row: CO(2--1) spectra for the central galaxy (left), northern tidal tail (center), and southern tidal tail (right), showing broad turbulent emission in the central galaxy and narrower profiles in the tails. Approximately 47\% of the CO luminosity is in the tidal tails.
}\label{fig:momentsSpecSDSSJ1448}
\end{figure*}

The data were jointly imaged for each object in \texttt{CASA} \citep{mcmullin_casa_2007} using natural image weighting to create CO(2--1) spectral cubes and continuum images. Natural weighting allows us to maximize sensitivity especially for low-surface-brightness emission at the cost of slightly lower spatial resolution. No significant continuum emission was detected in either dataset; thus, we did not perform continuum subtraction in the $uv$ or image plane. The final spectral cubes for both objects average four native channels, with spectral resolution of $\approx$34\,\kms and spectral line sensitivity of $\approx$90\,\uJybeam. The observations reach an angular resolution of 1.3\arc$\times$ 1.6\arc and  1.0\arc$\times$ 1.2\arc for \PSBI and \PSBII, respectively, with continuum sensitivity of 4.5\,\uJybeam for each.\footnote{The CO data for \PSBI were presented in \citetalias{spilker_star_2022}, which we reproduce here for completeness and to facilitate consistent analysis of both galaxies presented in this work.}   

\begin{figure*}[ht!]
\centering{
\includegraphics[width=0.99\textwidth]{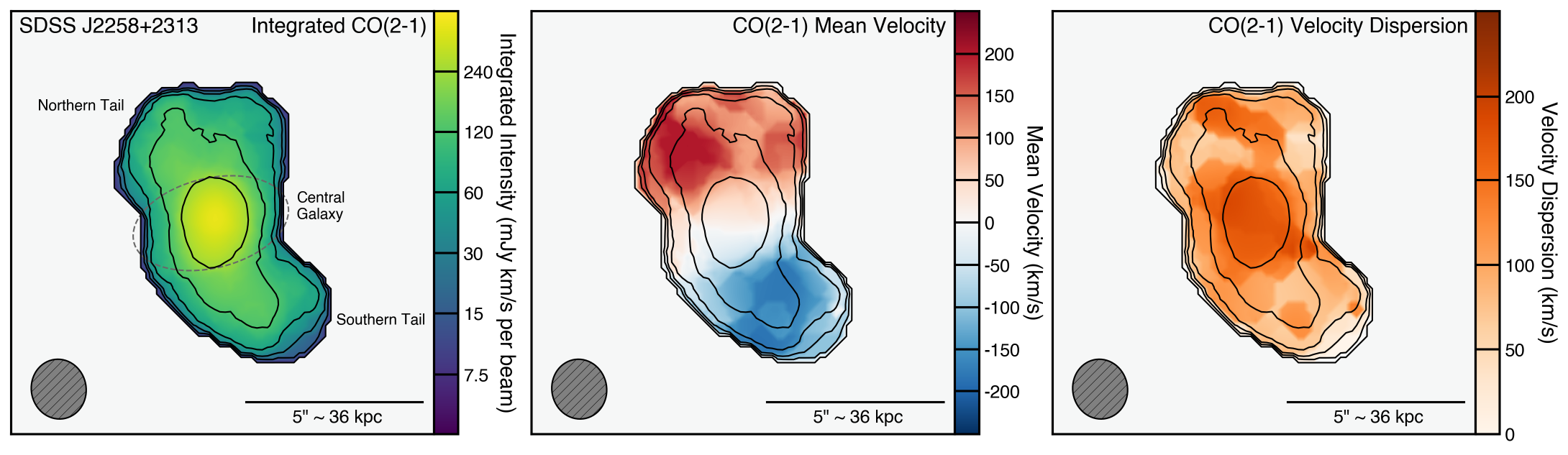}
\includegraphics[width=0.85\textwidth]{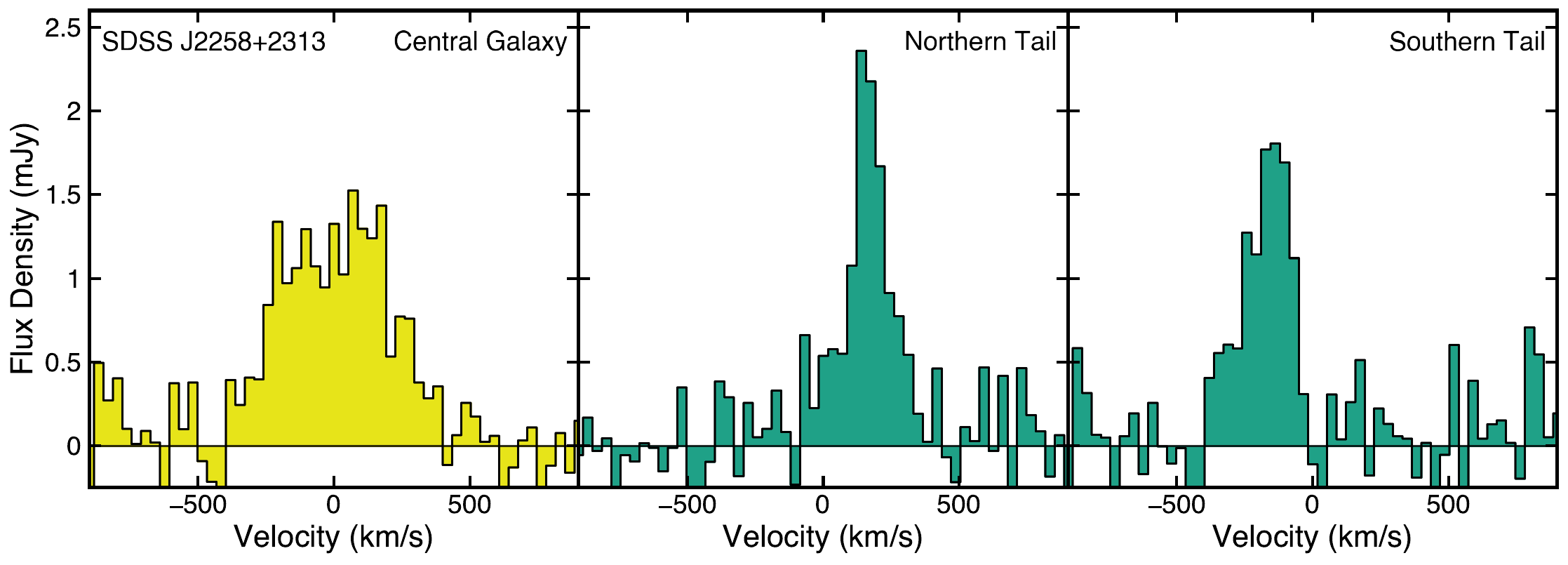}
}
\caption{Moment maps and spectra of ALMA CO(2--1) emission, as in Figure \ref{fig:momentsSpecSDSSJ1448}, but for \PSBII. CO emission extends up to $\sim$40 kpc, with redshifted northern and blueshifted southern tidal features. Approximately 55\% of the CO luminosity resides in the tidal tails.
}\label{fig:momentsSpecSDSSJ2258}
\end{figure*}

\begin{deluxetable*}{cccccccccc}
    \tablecaption{Multiwavelength properties for \PSBI and \PSBII. \label{tab:multiProps}}
    \tablecolumns{10}
    \tablewidth{0pt}
    \setlength{\tabcolsep}{2pt}
    \tablehead{
        \colhead{Object} & \colhead{Region} & \colhead{$L'_{\rm CO(2-1)}$}  & \colhead{$\rm M_{\rm H_2}$} & \colhead{$R_{\rm CO}$} & \colhead{$\sigma_{\rm CO}$} & \colhead{H$\alpha$ flux} & SFR & $S_{\rm6\,GHz}^{\rm int}$ & $S_{\rm6\,GHz}^{\rm peak}$ \\
        & & ($10^9$ K km s$^{-1}$ pc$^{2}$) & ($10^{10}$ \Msol) & (kpc) & (\kms) & ($10^{-16} \,\mathrm{erg/s/cm^2}$) & $(\Msol/\rm yr)$ & $(\mu \rm Jy)$ & $(\mu \rm Jy)$
    }
    \startdata
    & Central & $3.21\pm0.21$\tablenotemark{a} & $1.28\pm0.08$\tablenotemark{a} & $1.47\pm0.70$ & $282\pm21.0$ & $27.4_{-3.78}^{+4.31}$ & $22.4_{-3.09}^{+3.52}$\tablenotemark{b} & $97.9\pm2.03$ & $33.1\pm2.03$ \\
    & Northern (N1) & $0.56\pm0.21$ & $0.22\pm0.08$ & $2.17\pm0.77$ & $39.7\pm3.36$ & $<2.49$ & $<2.39$ & -- & --\\
    J1448+1010 & Northern (N2) & $0.44\pm0.21$ & $0.18\pm0.08$ & $<4.98$ & $<33.4$ & $3.27_{-2.95}^{+3.30}$ & $3.14_{-2.83}^{+3.16}$ & -- & -- \\
    & Northern (N3) & $0.91\pm0.21$ & $0.36\pm0.08$ & $<4.98$ & $<33.4$ & $2.93_{-2.71}^{+2.76}$ & $2.81_{-2.60}^{+2.65}$ & -- & -- \\
    & Southern & $1.04\pm0.21$ & $0.42\pm0.08$ & $0.98\pm0.84$ & $65.2\pm10.1$ & $<2.77$ & $<2.26$ & -- & -- \\
    \hline
    & Central & $4.03\pm0.23$ & $1.61\pm0.09$ & $1.75\pm0.73$ & $208\pm17.3$ & $5.24_{-1.01}^{+1.10}$ & $5.31_{-1.02}^{+1.11}$\tablenotemark{b} & $15.1\pm2.04$ & $14.8\pm2.04$\\
    J2258+2313 & Northern & $3.17\pm0.23$ & $1.27\pm0.09$ & $3.14\pm0.22$ & $75.4\pm7.58$ & $<3.04$ & $<3.08$ & -- & -- \\
    & Southern & $2.19\pm0.23$ & $0.88\pm0.09$ & $2.19\pm0.22$ & $80.9\pm8.71$ & $<2.50$ & $<2.53$ & -- & -- \\
    \enddata
    \tablenotetext{a}{The $L'_{\rm CO(2-1)}$ and $\rm M_{\rm H_2}$ measurements for the central galaxy of \PSBI are from \citetalias{spilker_star_2022}.}
    \tablenotetext{b}{The SFR measurement is likely to be a significant overestimate for \PSBI due to the presence of an AGN. This is potentially the case for \PSBII as well, but its AGN presence needs to be further studied.}
    \tablecomments{Limits for $R_{\rm CO}$ are the respective beam size, all other nondetections are \(3\sigma\).}
\end{deluxetable*}

\begin{figure}[ht!]
\centering{
\includegraphics[width=0.47\textwidth]{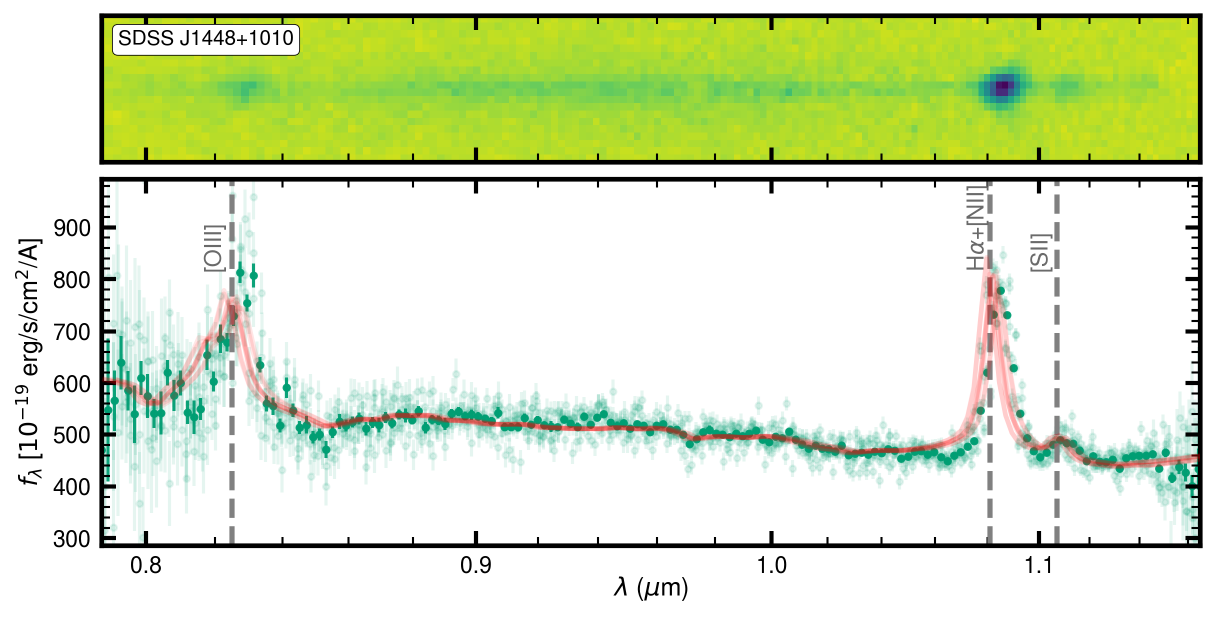}
}
\includegraphics[width=0.47\textwidth]{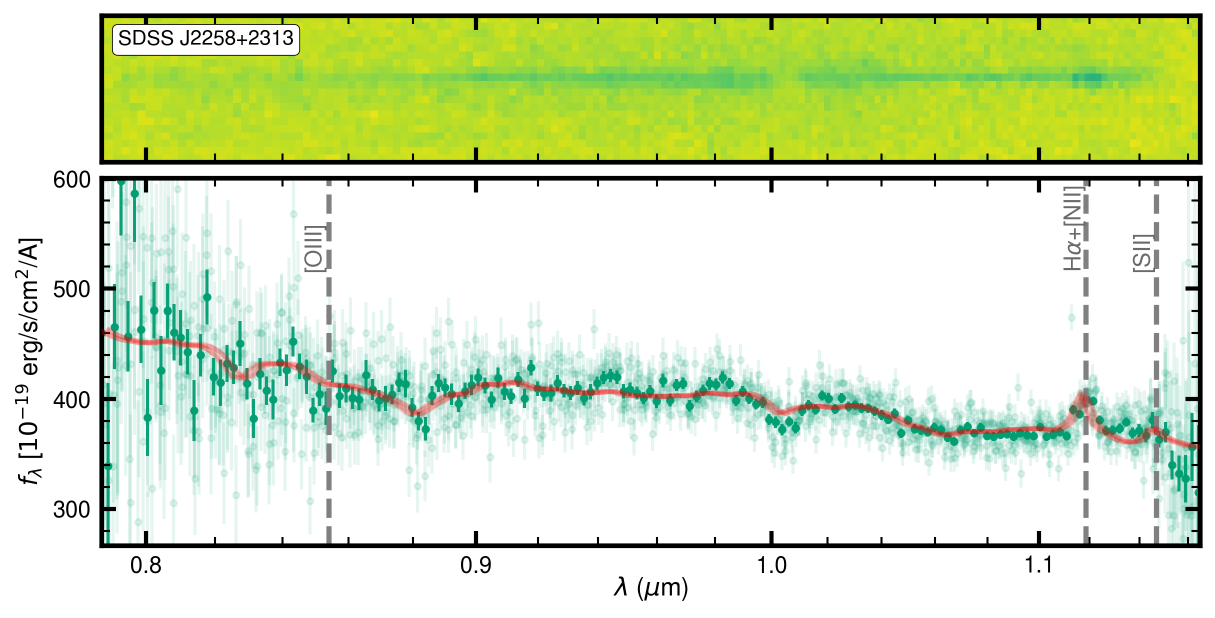}
\caption{HST WFC3/G102 grism spectra for \PSBI (top) and \PSBII (bottom). For each galaxy, the position angle-combined 2D grism spectrum is displayed above the 1D grism spectrum (green data points with error bars). The positions of [OIII], H$\alpha$+[NII], and the [SII] doublet are shown in gray dashed lines and the best-fit model is shown in the solid red line. 
}\label{fig:grismSpec}
\end{figure}

\begin{figure*}[ht!]
\centering
\hspace{0.0cm}
\includegraphics[width=0.96\textwidth]{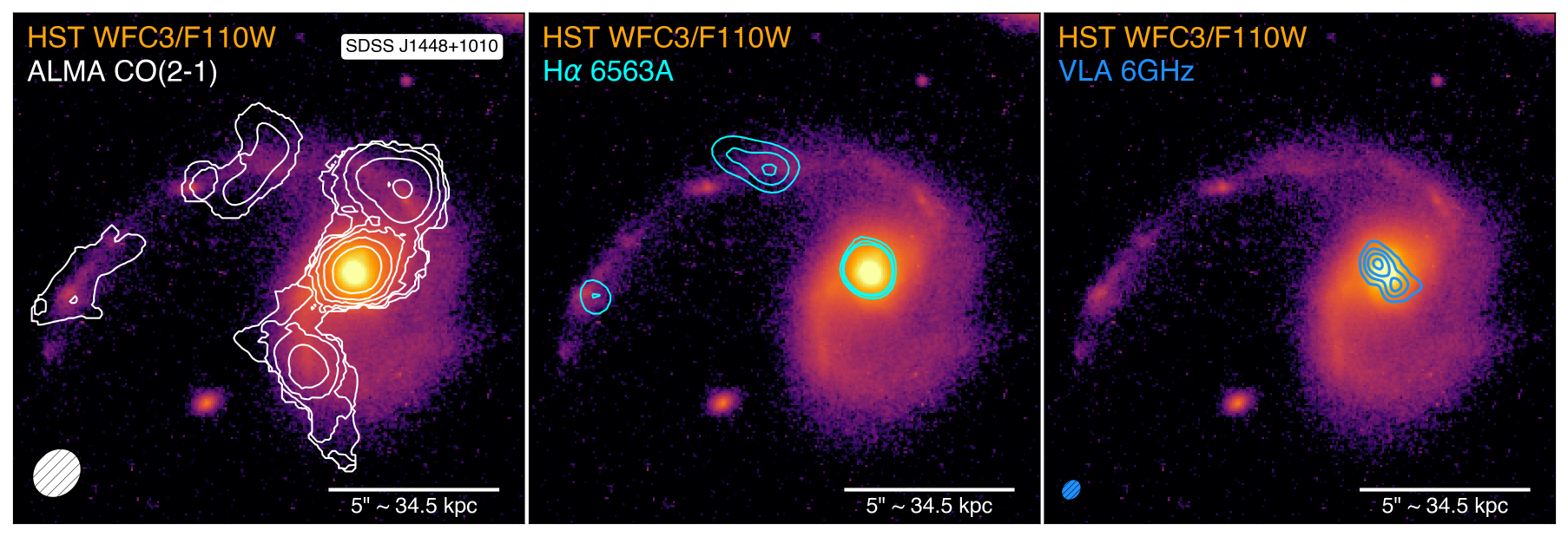}

\vspace*{-0.065in}
\hspace*{0.0cm}
\includegraphics[width=0.96\textwidth]{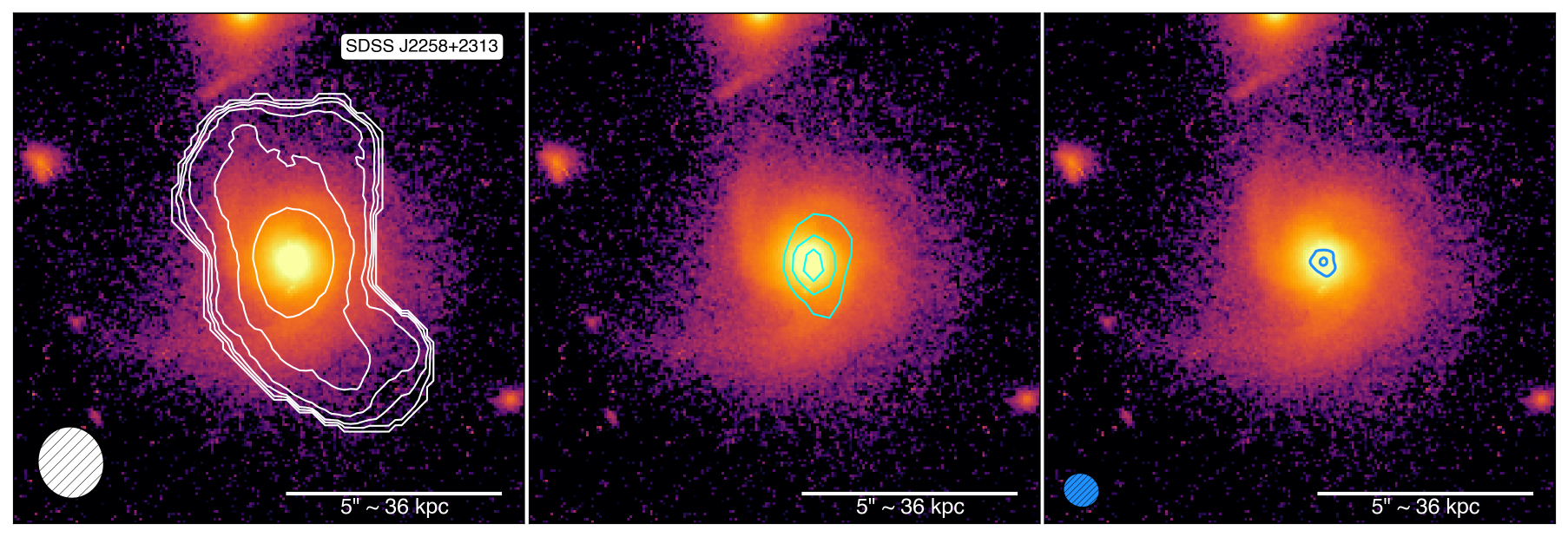}
\caption{Multiwavelength observations of the two PSBs, \PSBI (top row) and \PSBII (bottom row). The left column presents ALMA CO(2--1) contours (as in Figures \ref{fig:momentsSpecSDSSJ1448} and \ref{fig:momentsSpecSDSSJ2258}) highlighting the distribution of cold molecular gas. The ALMA synthesized beam is shown in the lower left corner. The middle column shows H$\alpha$ emission contours tracing regions of ongoing star formation, with contour levels drawn at 3, 6, and 9$\sigma$. The right column features VLA 6\,GHz radio contours drawn at 4, 8, 12, and 16$\sigma$ with the VLA synthesized beam shown in the lower left corner. All contours are overlaid on the HST F110W direct image for the respective galaxy. 
}\label{fig:multiSummary}
\end{figure*}

\subsection{HST Observations} \label{hstobs}
To investigate the structural properties and spatial distribution of the stellar light in our target PSBs, we obtained deep imaging with HST WFC3/IR using the F110W filter (PID 16201, PI Spilker). This band spans a wavelength range of 0.9-1.4\,$\mu$m which is well-suited for capturing the rest-frame optical light of these galaxies at their redshifts ($z\sim0.7$). The total integration time for each galaxy was 2.5\,ks which was divided into a 4 $\times$ 500\,s dither pattern and a final dither position with an exposure time of 450\,s. 

We also acquired G102 grism spectra for both galaxies, covering the wavelength range 0.8-1.15\,$\mu$m with a spectral resolution $R\sim210$ \citep{dressel_wide_2012}, which allow us to map the H$\alpha$ emission line. The G102 grism observations for both galaxies were conducted as part of program PID 16248 (PI Spilker). Shallow F098M direct images were also obtained during these visits which provided complementary rest-frame optical coverage. However, their shallow depth limits their sensitivity to the faint tidal tails, so we do not combine these with the deeper F110W images in our analysis. 

The WFC3 direct imaging and G102 grism exposures for both galaxies were reduced using the `Grism redshift \& line' analysis software \citep[\texttt{Grizli},][]{brammer_grizli_2019}, designed for the analysis of space-based slitless spectroscopic data. The raw WFC3 exposures were first processed through the standard HST pipeline, including bias subtraction, dark current correction, and flat-field correction to ensure accurate and uniform sensitivity across the detector. Following these preliminary corrections, the individual exposures were combined using the drizzle algorithm, implemented in \texttt{Grizli}, which resamples the images onto a finer grid with a final pixel scale of $0.06\arc$ / pixel. 

For the WFC3/G102 grism data, the reduction involved extracting two-dimensional (2D) spectra for each object from the raw exposures, applying similar corrections as done to the direct imaging. The \texttt{Grizli} pipeline then identified and extracted the spectra for each object in the field, accounting for the dispersion introduced by the grism. The pipeline then fits continuum and emission line models for the 2D spectra, thus isolating the blended H$\alpha$+[NII] emission lines from the underlying stellar continuum. Contamination-subtracted narrow band images centered on the H$\alpha$ emission were then created based on the known redshift of the target. These images were then drizzled to match the pixel scale of the direct images, ensuring consistency between imaging and spectroscopic datasets. 

H$\alpha$ emission maps were generated by subtracting the best-fit continuum model from the 2D grism spectra assuming that the direct image is representative of the morphology of the source. These maps were then drizzled to match the pixel scale of the direct images and are generated with dimensions of $80 \times 80$ pixels. The apertures used for the H$\alpha$ flux measurements were chosen to match the regions where CO emission was detected, ensuring consistency in spatial coverage between the H$\alpha$ and molecular gas measurements. The H$\alpha$ fluxes were then rescaled to account for contamination from [NII] by applying the scaling factor $f(\mathrm{[NII]/H\alpha})$ for each galaxy ($\sim$1.3 for \PSBI and $\sim$0.6 for \PSBII) as measured using Keck/NIRES in \citet{2025ApJ...981...60Z}.
We then estimate the star formation rates (SFRs) using the relation from \citet{kennicutt_star_1998}, corrected to a Chabrier IMF \citep{muzzin_well-sampled_2010}.

\subsection{VLA Observations} \label{vlaobs}
We observed both galaxies with the VLA C-band receiver in both the A-configuration and B-configuration. The B-configuration observations were conducted from 2020 July to 2020 October under project code 20A-188, while the A configuration observations were conducted in 2020 December under project code 20B-281. These combined configurations provide a more complete sampling of the uv-plane, allowing us to map the spatial distribution of the 6\,GHz emission with improved sensitivity and resolution. The observations used the standard C-band 3-bit continuum setup, providing continuous coverage from 4.0-8.0\,GHz in two 2\,GHz-wide basebands. The basebands were subdivided further into 128\,MHz subbands with 1\,MHz channelization. For \PSBI, the quasar 3C286 served as the primary flux and bandpass calibrator and J1445+0958 served as the phase calibrator. The primary flux and bandpass calibrator was the quasar 3C48 and J2254+2445 served as the phase calibrator for \PSBII. The phase calibrators for both targets were observed every six minutes in both configurations. The data from both configurations were reduced using the standard VLA pipeline, with additional manual flagging of radio frequency interference and other bad visibility data.

We combined the visibility data sets from both configurations to allow for improved imaging results. For both galaxies, imaging was performed with Briggs weighting \citep{briggs_high_1995} with a robust parameter of zero. For \PSBI, the resulting synthesized beam size is $0.46\arc\times0.37\arc$. The imaging process for \PSBII additionally required the use of the w-projection algorithm \citep{cornwell_noncoplanar_2008} to correct for non-coplanar baseline effects due to the wide field of view of the observations. This technique provided a synthesized beam size of $0.58\arc\times0.54\arc$. The use of the w-projection algorithm during the imaging process for \PSBI was not required as no bright, distant sources exist in the field of view of the observations. The final data cubes achieve rms noise levels of $\sim$2.0 $\mu$Jy beam$^{-1}$ for both galaxies.

\section{Analysis} \label{analysis}

\subsection{Molecular Gas Features in ALMA CO(2--1)} \label{CO21gas}
To assess the properties of the cold molecular gas features in our targets, we analyze the ALMA CO(2--1) emission observed in both the central galaxies and their extended tidal regions. Figures \ref{fig:momentsSpecSDSSJ1448} and \ref{fig:momentsSpecSDSSJ2258} show the CO(2--1) moment maps and spectra for \PSBI and \PSBII, respectively. The faint and narrow emission in the tidal regions necessitate a specialized method for creating the moment maps; see \citetalias{spilker_star_2022} for details on the procedure. In brief, we construct moment maps by identifying $\geq$4$\sigma$ peaks in the CO cube and dilating outward to the $2\sigma$ contour in each spectral channel. This approach avoids overwhelming the tidal features with noise from signal-free channels and ensures accurate flux measurements. The emission from the central galaxy is indicated by the dashed ellipse in the integrated CO maps for both objects, where the central component is separated from the northern and southern tidal tails by cutting perpendicular to the minimum level of the region between each feature. The CO(2--1) spectra include those for the central galaxy as well as the northern and southern tidal tail features.

The CO spectra of the central galaxy and extended features for each object were analyzed by fitting the spectra with single Gaussian components. The direct summation of the emission across channels with significant flux yields total line fluxes consistent with the values derived from the Gaussian fits. We used these fits to extract CO line widths, used as proxies for gas velocity dispersions. The spatial extent of each CO component was determined using the \texttt{CASA IMFIT} task on the integrated intensity maps to find their respective deconvolved sizes. Applying a standard CO-H2 conversion factor of \(\alpha_{\rm CO} = 4.0\) \(\rm M_{\odot}\)/ (K km s\(^{-1}\) pc\(^{2}\)), appropriate for non-starbursting, high-metallicity galaxies \citep[e.g.,][]{bolatto_co--h2_2013}, and assuming thermalized emission with $r_{21}=1.0$, we derived molecular gas masses for each component for each galaxy. The CO properties from the ALMA data for each post-starburst split into the central galaxy and northern (with subregions N1, N2, and N3 for \PSBI; see Fig. \ref{fig:momentsSpecSDSSJ1448}) and southern tails are included in Table \ref{tab:multiProps}. The tidal features contain approximately 47\% \(\pm\) 5\% of the total CO emission for \PSBI (\citetalias{spilker_star_2022}) and 55\% \(\pm\) 5\% for \PSBII.

\subsection{HST Grism Spectroscopic Characteristics} \label{HSTanalysis}
We focus on the spatial distribution of H$\alpha$ emission in the two PSBs via the HST grism spectra. In Figure \ref{fig:grismSpec}, we present the 1D and 2D HST WFC3/G102 grism spectra for our two PSBs. For \PSBI, the spectra exhibit prominent [OIII] and H$\alpha$ emission lines, which could indicate AGN activity (see \citet{2025ApJ...981...60Z} for further discussion on the line widths). In contrast, the spectra for \PSBII show only faint H$\alpha$ emission, but its origin - whether residual star formation or other ionization mechanisms - is less clear. The further implications of these features are discussed in Section~\ref{agn}.

Figure \ref{fig:multiSummary} presents the WFC3 imaging and the G102 grism H$\alpha$ maps (contours) of both target galaxies. Both galaxies exhibit prominent H$\alpha$ emission in their central regions, indicating ongoing star formation or AGN activity at the core. Notably, \PSBI also shows significant H$\alpha$ emission (upwards of $\sim5\sigma$) in the northern tidal tail, suggesting that star formation is taking place within the molecular gas in the tidal tails. \PSBII does not show H$\alpha$ emission outside the central region, highlighting a difference in the spatial distribution of potential star-forming activity between the two systems. The H$\alpha$ flux measurements for each region and their corresponding SFRs are included in Table \ref{tab:multiProps}.

\subsection{Radio Morphology via VLA Continuum} \label{AGNanalysis}
To investigate the radio properties of \PSBI and \PSBII we analyze the VLA 6\,GHz continuum observations, which provide insight into the presence of compact or extended radio emission structures in these galaxies. Figure \ref{fig:multiSummary} (right) shows the VLA 6\,GHz emission contours from the resulting images for both galaxies overlaid on the HST WFC3 direct images. The emission for \PSBI exhibits two clear peaks which reach a signal-to-noise ratio (S/N) of $\sim$20, which could be due to either compact or young radio jets/lobes. \PSBII shows a much fainter (S/N $\sim8$) continuum detection with a more compact structure. 

To further understand the nature of the radio morphology for each galaxy, we use the \texttt{CASA IMFIT} task to fit 2D Gaussian components to the images. The integrated $S_{\rm6\,GHz}^{\rm int}$ and peak $S_{\rm6\,GHz}^{\rm peak}$ flux densities are included in Table \ref{tab:multiProps}. The discrepancy between the peak and integrated flux densities for \PSBI indicates that the emission is not confined to a single point but spread over a larger area, consistent with the observed double-peaked morphology. Since \texttt{IMFIT} does not report reliable sizes for sources near or below the beam size, we use the analytic approximation for point sources from \citet{condon_errors_1997} for \PSBII. This results in a 2$\sigma$ upper limit of 0.15$''$ given the beam size and peak S/N, corresponding to a physical size of $<1.1$\,kpc. The similarity between the peak and integrated flux densities further suggests a relatively pointlike morphology.

\subsection{Cross-Checking Molecular Gas Mass: CO vs. Dust} \label{alphaco}
A precise estimation of the CO-H$_2$ conversion factor $\alphaco$ is particularly critical for accurately deriving molecular gas masses. We have assumed a Milky Way-like $\alphaco = 4.0$, typical of roughly solar metallicity clouds in virial equilibrium \citep{bolatto_co--h2_2013}. We use 2\,mm dust continuum measurements as an independent method to cross-check our $\alphaco$ assumptions, offering a test for consistency. Weak continuum emission $S_{142\,\rm GHz}=21.1\pm4.7 \, \mu\rm Jy$ is detected for the central galaxy of \PSBII. The flux ratio between the upper (149\,GHz) and lower (137\,GHz) sidebands is $1.3\pm0.4$, consistent with the expectation for dust emission on the Rayleigh-Jeans tail and suggesting no significant contribution from synchrotron emission. We estimate a dust-based molecular gas mass of $M_{\rm mol,gal}=(1.6\pm0.4)\times10^{10}\,\Msol$ for the central galaxy of \PSBII, using standard dust emissivity assumptions and a mass-weighted dust temperature of 25\,K \citep[e.g,][]{dunne_census_2003,scoville_ism_2016,privon_interpretation_2018,liang_dust_2019}. This implies a CO-H$_2$ conversion factor of $\alphaco \approx 3.9$ which is in agreement with our fiducial assumptions; a similar conclusion was reached for \PSBI (\citetalias{spilker_star_2022}). 

For the tidal tails in both PSBs we do not detect dust continuum emission at 2\,mm and thus report $3\sigma$ upper limits. The spatial extent and faint nature of the tails make such detections challenging given the sensitivity limits of our observations. For \PSBI the northern tail has a dust-based upper limit of $\rm M_{\rm mol,tail}<3.3\times10^{10}\,\Msol$ while the southern tail has $\rm M_{\rm mol,tail}<2.6\times10^{10}\,\Msol$. The limits are $\rm M_{\rm mol,tail}<5.9\times10^{10}\,\Msol$ and $\rm M_{\rm mol,tail}<5.5\times10^{10}\,\Msol$ for the northern and southern tails of \PSBII, respectively. These limits are consistent with our assumptions of $\alphaco$. 

Although we presently have no evidence for (or against) this scenario, it is possible that $\alphaco$ differs between the tidal tails and the central galaxies. If we adopt a Milky Way-like $\alphaco = 4.0$ for the central galaxies but a ULIRG-like $\alphaco = 1.0$ in the tails, the inferred molecular gas fractions in the tails would be significantly lower, with their total contributions decreasing to approximately 20\% for \PSBI and 25\% for \PSBII. On the other hand, if we assume the opposite - a lower $\alphaco$ in the galaxies and a higher value in the tails - the inferred gas masses in the tails would increase substantially, illustrating the sensitivity of our conclusions about gas distribution to the adopted $\alphaco$ values.

\subsection{Evidence of AGN} \label{agn}

We utilize our multiwavelength observations to investigate the presence of AGN in these two PSBs, which is critical for understanding the mechanisms driving star formation quenching. The existence of an optical AGN in \PSBI has already been suggested based on its [OIII]/H$\beta$ line ratio in \citet{greene_role_2020}, with \PSBI exhibiting by far the most luminous and broadest [OIII] emission in the sample, and further speculated upon in \citetalias{spilker_star_2022}. \PSBI shows clear signs of AGN activity across multiple wavelengths presented in this work. VLA 6\,GHz observations (Section~\ref{vlaobs}, Figure \ref{fig:multiSummary}) show a radio morphology consistent with compact/young radio jets, a characteristic signature of AGN activity. The HST WFC3/G102 grism spectrum in Figure \ref{fig:grismSpec} exhibits broad and prominent emission lines, including strong [OIII] and H$\alpha$ emission. These observations are consistent with the results from Keck/NIRES and SDSS observations which confirm the presence of strong optical emission lines in \PSBI, with their classification via emission line diagnostics further suggesting \PSBI is AGN dominated \citep{2025ApJ...981...60Z}. 

\begin{figure}[ht!]
\centering{
\includegraphics[width=0.48\textwidth]{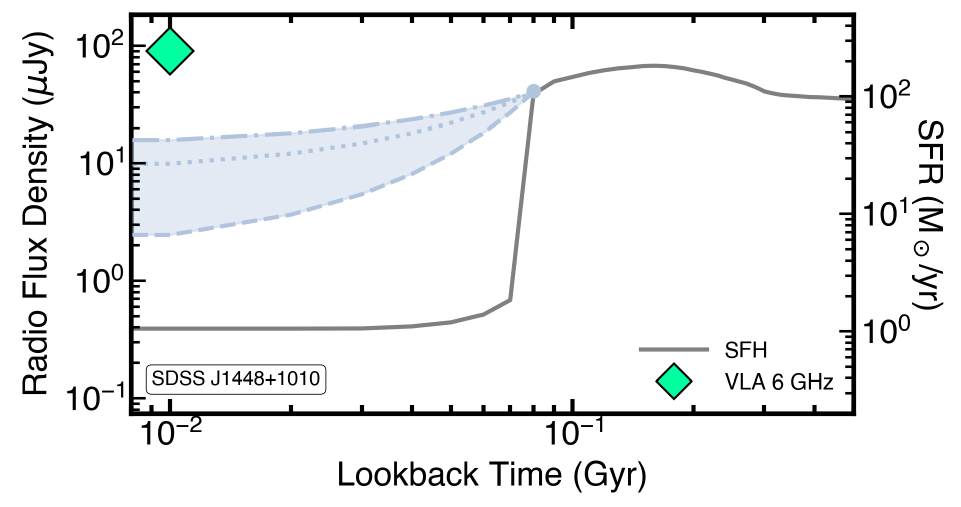}
}
\includegraphics[width=0.48\textwidth]{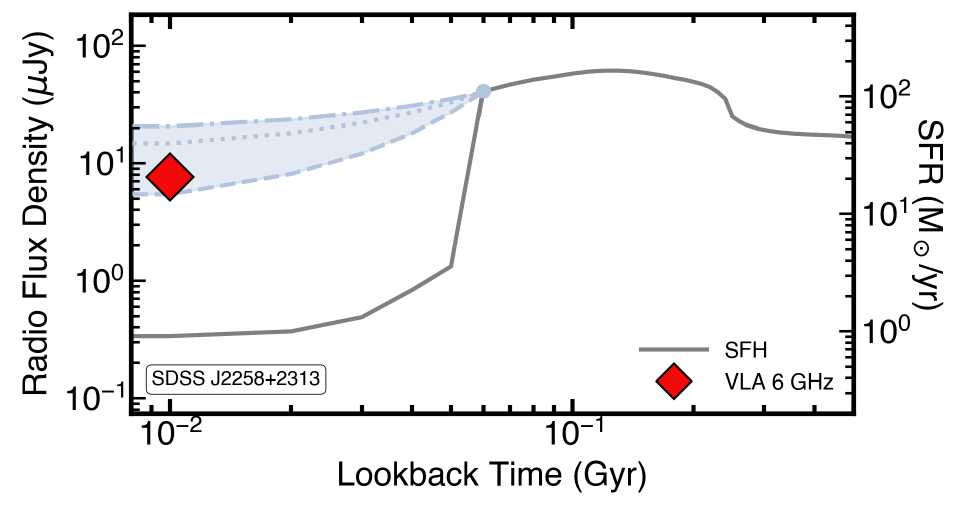}
\caption{Predicted versus observed VLA radio flux densities for \PSBI (top) and \PSBII (bottom). The solid gray line shows each galaxy's reconstructed SFH, derived from SED fitting of the SDSS spectrum and photometery \citep{suess_mathrmsquiggecle_2022}, with the observed radio flux (diamond) in green and red, respectively. The predicted radio decay curve (shaded band) is computed assuming an exponential decline from the SFR at the onset of quenching for multiple fiducial radio star formation timescales: 50\,Myr (dashed), 100\,Myr (dotted), and 150\,Myr (dash-dotted). \PSBI's radio flux exceeds predictions, indicating AGN dominance, while \PSBII's emission aligns with residual star formation.
}\label{fig:radioPredicted}
\end{figure}

The evidence for AGN activity is inconclusive for \PSBII. \PSBII shows only faint H$\alpha$ emission, making it difficult to identify AGN activity from the WFC3/G102 data alone. The radio emission is compact and weak, making it less conclusive for AGN identification, and thus does not exhibit clear signs of AGN activity based on morphology. In fact, the radio emission is sufficiently weak that it may be due to past star formation in this recently-quenched galaxy. 

We investigate the radio emission by comparing the present-day radio flux density and the predicted radio flux density -- the case where the radio comes from past star formation -- based on the star formation histories (SFHs) of both galaxies. Radio continuum at 6\,GHz arises mainly from non-thermal synchrotron emission from cosmic rays accelerated by supernova remnants, which trace star formation over timescales of up to several hundred Myr. For \PSBII, the observed radio emission is weak enough to raise the possibility that it originates from star formation. Using a standard conversion between radio luminosity and SFR \citep{murphy_calibrating_2011}, we find that the SFR derived from the observed radio luminosity ($\sim$20\,\sfr) exceeds the `instantaneous' SFR derived from spectral energy distribution (SED) fitting ($\sim$1\,\sfr, \citealt{bezanson_now_2022}). This discrepancy could be explained by the long timescale over which radio synchrotron emission traces past star formation. 

To evaluate whether the observed radio emission could plausibly originate from past star formation in \PSBII, we consider a simplified model where the radio luminosity fades exponentially following the suppression of star formation. The radio luminosity at a time $t$ is given by,
\begin{equation} \label{eq:1}
    L_\mathrm{radio,t} = L_\mathrm{radio,0}\,e^{-t/\tau},
\end{equation}
where $L_\mathrm{radio,0}$ is the initial radio luminosity and $\tau$ is the characteristic decline timescale. While the detailed shape and timescale of the decay depend on the injection and cooling of relativistic electrons, previous studies have shown that declines on tens to hundreds Myr timescales are broadly consistent with expectations for synchrotron and inverse Compton cooling \citep[e.g.,][]{condon_radio_1992,murphy_far-infrared-radio_2009, arango-toro_probing_2023}. We therefore calculate predicted radio flux densities for three fiducial values of $\tau$ (50, 100, and 150\,Myr) and consider an assumed exponential decline computed from the SFR at the onset of quenching. Thus, Eq. \ref{eq:1} provides a simplified prescription of the evolution of radio luminosity as opposed to a precise modeling of the physical onset of radio fading.


Figure \ref{fig:radioPredicted} presents the comparison between the observed VLA 6\,GHz flux density and the predicted radio flux density derived from the SFH of each post-starburst. For \PSBI, the observed radio flux density is significantly higher than any reasonable prediction for star formation, even under the assumption of long synchrotron timescales. This result strongly supports the conclusion that the radio emission in \PSBI is dominated by AGN activity. In contrast, the radio emission from \PSBII aligns with predictions for past star formation, indicating that it could plausibly originate from residual star formation rather than AGN activity. While this analysis provides valuable insights, we note that the results are not intended to be definitive, as the actual timescale over which radio synchrotron emission traces star formation remains uncertain.

\section{Results: Characterization of Extended Tidal Features} \label{resultstails}

\subsection{Are These Extended Features Tidal Tails?} \label{tidaltails}
Tidal tails are a natural consequence of strong gravitational interactions, particularly in major mergers, where tidal forces can remove material from the progenitor galaxies and redistribute it into extended features. For \PSBI, \citetalias{spilker_star_2022} argued that its extended features were tidal in nature based on the close spatial coincidence between the extended molecular gas and stellar tidal features, characteristic of tidal tails produced in gas-rich mergers. The presence of a secondary stellar peak near the galaxy center - perhaps evidence of the remnant of a merging companion (see Fig. 3 of \citetalias{spilker_star_2022}) - also supports the interpretation of a recent interaction, consistent with features commonly seen in post-merger systems.

We apply a similar analysis to \PSBII, where in Appendix \ref{J2258Residuals} we present the original WFC3 F110W image, a best-fit single S\'{e}rsic model of the stellar light, and the residual image after subtraction of the model. The residual image highlights a minor peak in the stellar light as a clump at the base of the northern tail, perhaps evidence of the remnant of a merging companion. Additionally, the large-scale asymmetry of the extended features suggests that they are not simply part of a dynamically stable disk, but rather tidal features formed through gravitational interaction. Perhaps the best evidence of an interaction, though, is the presence of a peak in the stellar light at the tip of the southern tail. This peak represents a possible tidal dwarf galaxy (TDG), which can form in major mergers as mass streams along tidal tails and collects at the tips (e.g., \citealt{duc_young_1998,duc_formation_2000}; see Section~\ref{sfprops} for further discussion). Although other nearby galaxies are visible in each field, the new HST grism data and existing SDSS photometric/spectroscopic data do not confirm redshifts for any of these sources. Thus, we do not know if they are at the same redshift as the respective post-starburst. While the evidence for a major merger is clearly less strong in \PSBII than \PSBI, taken together, the possible TDG, asymmetric arms, and secondary peak near the galaxy center all point to a past gravitational interaction.

\subsection{Did Tidal Gas Removal Have a Role in Quenching?} \label{molgasfeatures}

\begin{figure*}[ht!]
\centering
\includegraphics[width=0.9\textwidth]{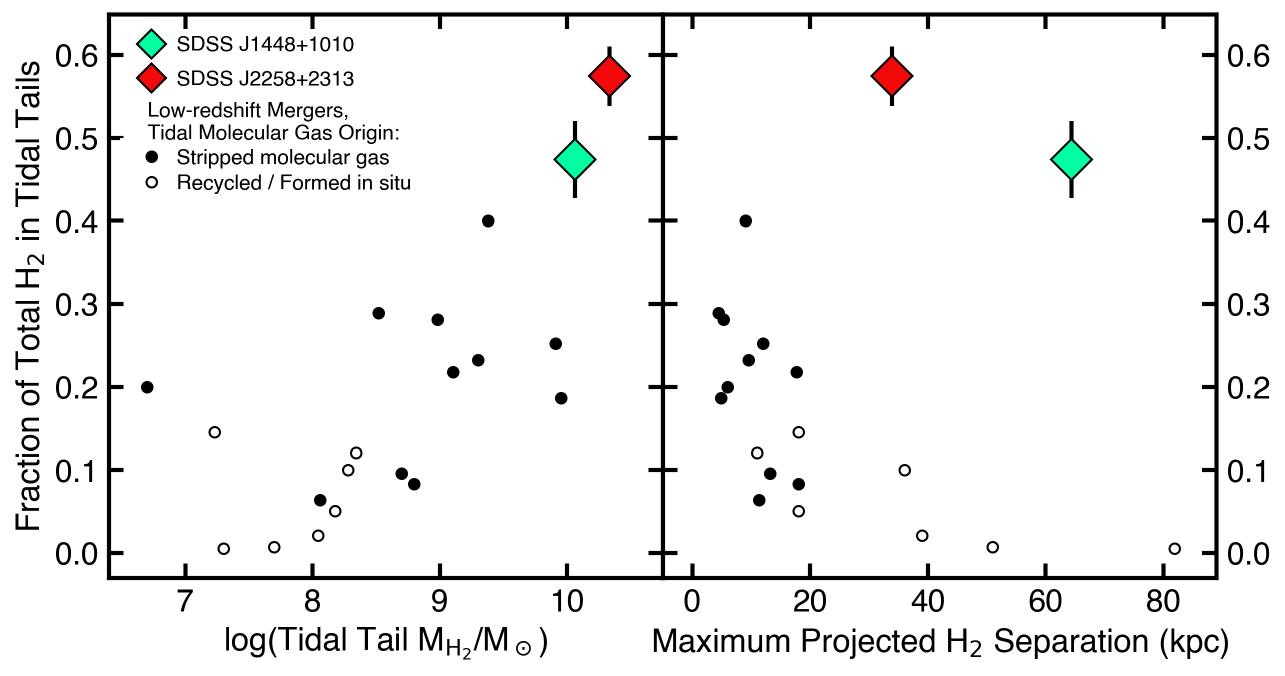}
\caption{Molecular gas properties of tidal tails in \PSBI (green) and \PSBII (red) compared to a sample of low-redshift mergers. The fraction of H$_2$ located in the tidal tails is plotted against the total gas mass in the tails (left panel) and the maximum projected separation of H$_2$ from the central galaxy (right panel). The comparison sample comprises 18 low-redshift ($z \leq 0.3$) galaxy mergers with detected CO in their tidal tails. This sample is further divided based on the origin of the molecular gas: either stripped from the host galaxy (black) or recycled or formed in situ (white). 
}\label{fig:tailsGasProps}
\end{figure*}

Both galaxies show evidence of interactions in the recent past, inferred from their disturbed morphologies in both HST and ALMA maps, which were likely responsible for removing much of the molecular gas into the extended features we observe. In Figure \ref{fig:tailsGasProps}, we analyze the molecular gas properties in the tidal tails of \PSBI and \PSBII, using the comparison sample assembled in \citetalias{spilker_star_2022}. The left panel shows the fraction of H$_2$ in the tidal tails relative to the entire galaxy compared to the total gas mass of the tails, while the right panel depicts the fraction of H$_2$ in the tails against the maximum projected separation of the H$_2$ gas from the galaxy. 

The tidal gas in \PSBI and \PSBII is highly unusual in both magnitude and spatial extent. Figure \ref{fig:tailsGasProps} shows that the gas features in the tidal tails are extreme when compared to the nearby merger sample. In both galaxies, approximately half of the molecular gas is located in the tidal features, significantly higher than in the low-redshift mergers. The maximum extent of the detected CO emission in the two PSBs is also extreme, especially given the large amount of gas in the tidal tails, with distances of $\sim$64 and $\sim$34\,kpc for \PSBI and \PSBII, respectively. While the alignment between the stellar and gas tidal features in \PSBI suggests tidal forces are responsible for removing the molecular gas in this system, the misalignment in \PSBII may reflect differences in the merger geometry or initial conditions of the interacting galaxies, potentially leading to the distinct spatial distributions observed. Variations in gas dissipation could also contribute to the distinct spatial distributions found, as observed for the tidal tails of a number of nearby systems \citep[e.g.,][]{smith_atomic_1997,hibbard_180_1999,hibbard_neutral_2000}.

Taking into account the difficulty of prescribing an exact quenching mechanism to a system, it is important to also consider alternative physical processes. One possibility is that AGN feedback contributed to the initial suppression of star formation (as opposed to the maintenance of the suppression as discussed in Section~\ref{lowSFRs}). In both galaxies, the scale and magnitude of the molecular gas displaced via tidal removal are much larger than what is typically observed in AGN-driven outflows, which generally eject only 1-10\% of the cold gas mass at high velocities \citep[e.g.,][]{fluetsch_cold_2019,spilker_ubiquitous_2020}, compared to the more gradual mechanism ($\sim$Gyr timescales) offered by the tidal removal of gas. Mechanisms such as virial shock heating \citep[e.g.,][]{keres_how_2005,dekel_galaxy_2006} or environmental effects \citep[e.g.,][]{sobral_dependence_2011,jachym_abundant_2014,darvish_effects_2016}, are also unlikely to be primary drivers. These tend to operate more efficiently in dense environments, whereas both PSBs are isolated galaxies undergoing major mergers. As discussed in Section \ref{tidaltails}, none of the potential nearby companions have thus far been confirmed to lie at the same redshift as the PSBs. We also find no evidence that the molecular gas has preferentially been stripped to one side in either of the galaxies, as expected in an environmental scenario like ram-pressure stripping \citep{jachym_abundant_2014}. While multiple processes could contribute a secondary role to the quenching of star formation in these PSBs, the tidal removal of cold gas during a major merger is most likely to be the driving mechanism of quenching in these PSBs.


\begin{figure}[ht!]
\centering
\includegraphics[width=0.48\textwidth]{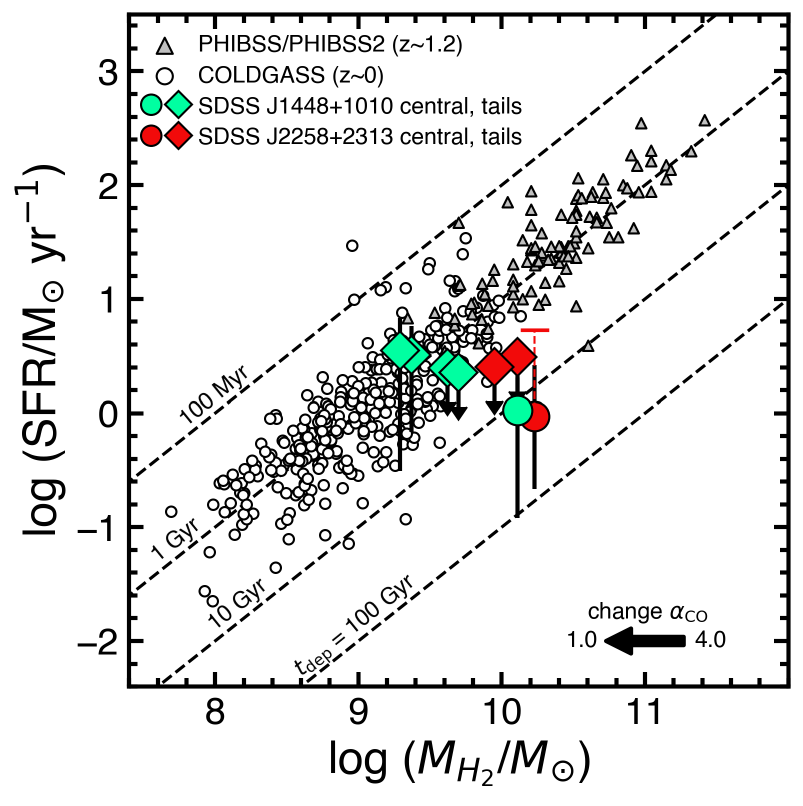}
\caption{SFR vs H$_2$ gas mass for the two post-starburst galaxies \PSBI and \PSBII. The central galaxies of \PSBI and \PSBII are shown as green and red circles (using their SED-derived SFRs), respectively, with their corresponding tidal tail components plotted as diamonds. Regions not detected in H$\alpha$ are shown as upper limits and are $3\sigma$. For comparison, we include star-forming galaxies at $\langle z \rangle \sim1.2$ from PHIBSS/PHIBSS2 \citep[gray triangles,][]{tacconi_phibss_2013} and massive galaxies at $z \sim 0$ from COLDGASS \citep[white circles,][]{saintonge_cold_2011-1}. Black dashed lines represent various constant gas depletion timescales ($\tau_\mathrm{dep} \equiv \mathrm{M}_{\mathrm{H_2}}/\mathrm{SFR}$). The red dashed line indicates the range between the SED-derived and Ha-inferred SFRs for the central of \PSBII. Although the central galaxies exhibit significantly longer gas depletion times than expected based on their gas reservoirs, two clumps in the northern tidal tail of \PSBI have depletion times typical of normal star-forming galaxies. 
}\label{fig:ksplot}
\end{figure}

\subsection{Is Star Formation Ongoing in the Tails?} \label{sfprops}

The star-forming properties of both the central galaxies and their extended tidal features provide insights into the quenching mechanisms at play in these systems. To assess how efficiently gas is converted into stars across the regions in these PSBs, we show SFR vs $\rm M_{\rm H_2}$ in Figure \ref{fig:ksplot}. The SFRs derived from SED-fitting \citep{bezanson_now_2022}, rather than H$\alpha$ emission, are used for the central galaxies due to concerns that the H$\alpha$ emission may be contaminated by AGN activity (see Section~\ref{agn}; \citealt{2025ApJ...981...60Z}). However, for reference, we indicate the H$\alpha$-inferred SFR for the central galaxy of \PSBII as a dashed line extending from the SED-derived value since the case of its H$\alpha$ emission arising from star formation has not been ruled out. To place these PSBs in a broader context, we compare them to a sample of star-forming galaxies at $\langle z \rangle = 1.2$ from PHIBSS/PHIBSS2 surveys \citep{tacconi_phibss_2013} and massive galaxies at $z \sim 0$ from COLDGASS \citep{saintonge_cold_2011-1}. Depletion timescales ($\tau_\mathrm{dep} \equiv \mathrm{M}_{\mathrm{H_2}}/\mathrm{SFR}$) ranging from 100\,Myr to 100\,Gyr are shown in black dashed lines.

The tidal tails of both galaxies exhibit varying star-forming properties. In the case of \PSBI, H$\alpha$ detections in regions of the northern tidal tail reveal star formation efficiencies (SFEs) similar to those of normal star-forming galaxies. This suggests that despite being stripped from the central galaxy, gas in this region is still capable of forming stars efficiently, which may indicate that some of the gas in the tidal tails has not been heated or disrupted to the extent that would prevent star formation. The fact that the star formation in this region is consistent with normal galaxies highlights the potential for tidally removed gas to contribute to star formation even in post-quenching systems. In contrast, the other tidal components of \PSBI, as well as those in \PSBII, are not detected in H$\alpha$ and are plotted as upper limits ($3\sigma$) in Figure \ref{fig:ksplot}. These non-detections suggest that while molecular gas is readily available in these regions, it is not actively forming stars.

The detection of H$\alpha$ emission in the extended tail regions of \PSBI opens up the question of the long-term fate of these features. One possibility is that these star-forming regions could eventually contribute to the formation of extended stellar halos around the host galaxy. Extended stellar halos, often observed in nearby massive galaxies, are thought to result from the accretion of satellite galaxies and the redistribution of stars during merger events \citep[e.g.,][]{tal_frequency_2009, duc_atlas3d_2015}. Another potential outcome is the formation of TDGs, which are observed in many $z\sim0$ mergers. In this scenario, gas in the tidal tails becomes self-gravitating, and densities can become high enough to form a dwarf galaxy that contains very little dark matter. If the stars forming in the tidal tails survive long enough to become gravitationally bound to the galaxy, they could form distinct TDG structures over time. This scenario is supported by observations of tidal debris around massive galaxies, which frequently show signs of star formation within them \citep[e.g.,][]{duc_young_1998, braine_formation_2000, duc_formation_2000}.


\begin{figure*}[ht!]
\centering
\includegraphics[width=0.4499\textwidth, trim={0 0 0 0cm}]{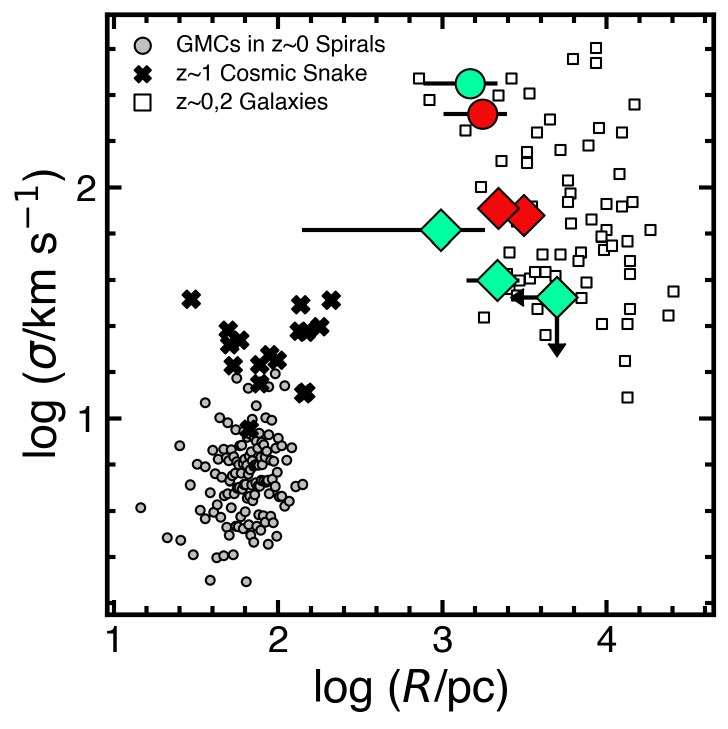}
\includegraphics[width=0.4499\textwidth, trim={0 0 0 0cm}]{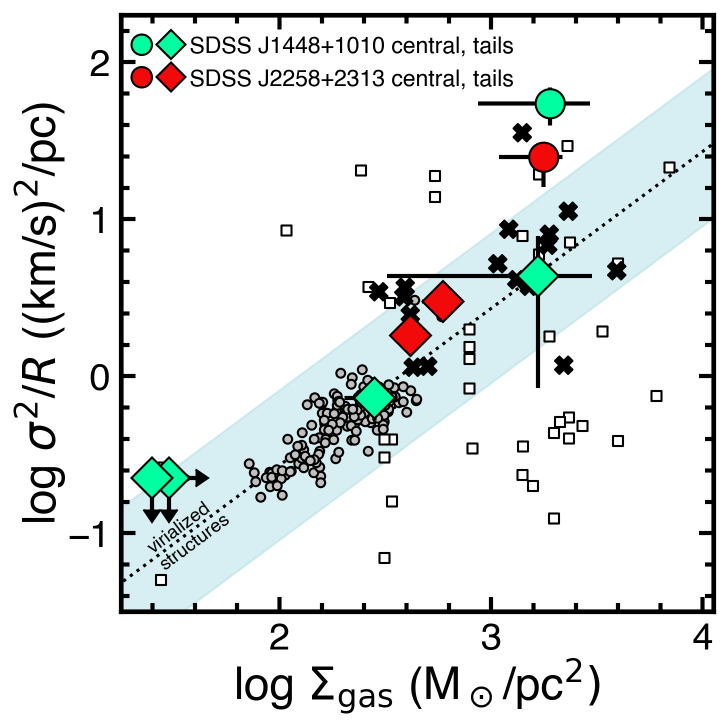}
\caption{Turbulence properties of the molecular gas in \PSBI and \PSBII. The left panel shows the relationship between size and velocity dispersion, while the right panel presents the dispersion normalized by size against the gas surface density. Markers for \PSBI and \PSBII are the same as in Figure \ref{fig:ksplot}. We include for comparison GMCs in $z\sim0$ spirals from PHANGS-ALMA \citep[gray circles;][]{rosolowsky_giant_2021}, clumps from the lensed $z\sim1$ `Cosmic Snake' galaxy \citep[black crosses;][]{dessauges-zavadsky_molecular_2019}, and both nearby and $\langle z \rangle \sim2$ `normal' galaxies \citep[white squares;][]{genzel_study_2010, leroy_molecular_2013, tacconi_phibss_2013}. While the tidal tails of both PSBs exhibit characteristics consistent with gravitationally bound structures, the central galaxies show significantly enhanced turbulence likely contributing to their suppressed star formation. 
}\label{fig:turb}
\end{figure*}

\section{Results: Star Formation Suppression in the Central Galaxies} \label{resultscentrals}

\subsection{Are The Central Galaxies Forming Stars?} \label{actuallyquenched}
The massive amount of molecular gas retained in these PSBs raises questions about whether they are truly quenched, as such gas reservoirs are typically expected to fuel ongoing star formation. The central galaxies of both \PSBI and \PSBII appear inefficient in their ability to form stars given their significant molecular gas reservoirs (Figure \ref{fig:ksplot}). Both central galaxies exhibit depletion timescales $\sim$10\,Gyr, much longer than typical star-forming galaxies. This very inefficient star formation appears to be a common feature of young PSBs in the \squiggle sample (\citealt{bezanson_now_2022}; Setton et al. in prep.).\footnote{\citet{bezanson_now_2022} consider the central galaxies and extended gas of \PSBI and \PSBII as single components, while our new data allows us to separate these features. The depletion times remain unusually long even when attributing $\sim$50\% of the total cold gas to the tidal features.} 

The optical spectra of \squiggle PSBs display strong Balmer absorption lines indicative of A-type stars, suggesting that significant star formation dramatically decreased within the past $\sim$1\,Gyr. Although not the case for \PSBI and \PSBII, most \squiggle galaxies even show H$\alpha$ in absorption, indicative of a lack of active, unobscured star formation \citep{2025ApJ...981...60Z}. In \PSBI, the observed H$\alpha$ and [OIII] emission lines are significantly broader than expected ($\sim$30\,\rm \AA) for star-forming regions, suggesting a significant AGN component. The 6\,GHz radio continuum aligns with predictions for residual star formation for \PSBII (see Section~\ref{agn}; \citealt{2025ApJ...981...60Z}).

For \PSBI, the mid-J CO(4-3) line shown in Appendix \ref{CO43} further complicates this picture (no other CO lines are available for \PSBII). The ratio in flux units for the CO(4-3) and CO(2-1) lines r$_{42}=2.43\pm0.13$ is higher than typical values for recently quenched galaxies or the inner disk of the Milky Way \citep[e.g.,][]{fixsen_cobe_1999}, but lower than those observed in extreme star-forming galaxies or AGN \citep[e.g.,][]{spilker_rest-frame_2014, harrington_turbulent_2021}. This suggests that \PSBI must contain sufficient warm and/or dense gas to produce this line ratio. However, whether this gas is primarily associated with residual star formation or is heated by AGN activity remains unclear. We are pursuing additional mid-J CO observations in ongoing work. To better probe obscured star formation, using tracers such as Pa$\alpha$ or mid-IR fine-structure lines with JWST will be crucial as these tracers are less affected by dust attenuation. Taken together, while we find strong evidence that star formation has declined sharply in both galaxies, we cannot definitely rule out low levels of residual star formation, particularly in \PSBII.

\subsection{What is Suppressing Star Formation in the Central Galaxies?} \label{lowSFRs}
We have shown that roughly half of the total molecular gas in each system exists in the tails, but the other half still remains in the central galaxies (subject to assumptions about $\alpha_{\rm CO}$). In this section, we aim to understand why the central galaxies are forming stars inefficiently given their massive gas reservoirs by investigating the physical state of the molecular gas as well as the role of potential feedback from radio AGN.

\subsubsection{Turbulence in the Molecular Gas} \label{phystate}

\begin{figure*}[ht!]
\centering
\includegraphics[width=0.9\textwidth]{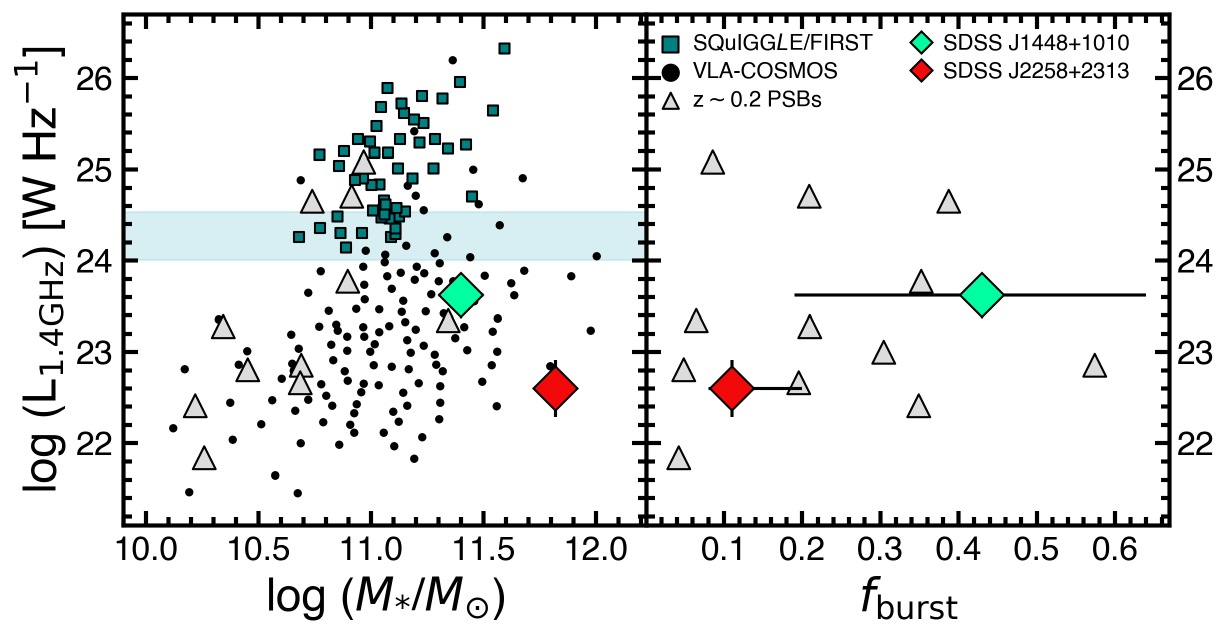}
\caption{The left panel shows the 1.4\,GHz radio luminosity plotted against the total stellar mass. Shown alongside are comparison samples including radio AGN in \squiggle galaxies detected in FIRST \citep[green squares;][]{greene_role_2020}, VLA-COSMOS $z\sim1$ AGN \citep[black circles;][]{smolcic_cosmic_2009}, and $z\sim0.2$ PSBs with radio AGN \citep[gray triangles;][]{shin_radio_2011}. The radio detection limit from \squiggle/FIRST is indicated with the shaded band and corresponds to a luminosity limit range $L_{1.4\rm GHz} = (1.0-3.4)\times 10^{24}\,\rm{W\,Hz^{-1}}$. The right panel plots the 1.4\,GHz radio luminosity against the fraction of stellar mass associated with the starburst $f_{\rm burst}$.
}\label{fig:LradioMstar}
\end{figure*}

We investigate the turbulence properties of the molecular gas in these PSBs. While the physical scales we probe ($>$1\,kpc) differ from the smaller sub-kpc scales typically considered for giant molecular clouds (GMCs), it is instructive to apply similar turbulence-scaling laws to understand how gas dynamics regulate star formation in these larger regions. It is not uncommon to see similar discussions that include galaxy-integrated measurements in the literature \citep{nguyen-luong_scaling_2016, salim_spinning_2020}, applied even to PSBs \citep{smercina_after_2022}. The ISM turbulent pressure applied to low-z systems in \citet{smercina_after_2022} is proportional to the size and velocity dispersion quantities we apply, just parameterized in a different manner. Throughout this section we use the velocity dispersions obtained from single-Gaussian fits to the CO(2--1) line profiles for each region. We do not correct these dispersions for ordered motions and therefore interpret them as proxies for the intrinsic gas dispersion.    


In the left panel of Figure \ref{fig:turb}, we examine the size-velocity dispersion relation for the central galaxies and their associated tidal tail clumps, using the quantities derived from the integrated CO maps reported in Table \ref{tab:multiProps}. We compare to $\sim$150\,pc resolution analysis of GMCs in $z\sim0$ spirals from PHANGS-ALMA \citep{rosolowsky_giant_2021}, clumps from the lensed $z\sim1$ `Cosmic Snake' galaxy \citep{dessauges-zavadsky_molecular_2019}, and both nearby and $\langle z \rangle \sim2$ `normal' galaxies from the literature \citep{genzel_study_2010, leroy_molecular_2013, tacconi_phibss_2013}. In comparison to the other galaxy-integrated measurements from the literature, both \PSBI and \PSBII show evidence of increased turbulence. The velocity dispersions of $\sim$200\,\kms on $\sim$1\,kpc scales are higher than $\sim$90\% of the comparison galaxies. The smaller-scale tidal tail regions also appear to have moderately enhanced velocity dispersions in comparison to GMCs in nearby galaxies or the `Cosmic Snake', although this is more qualitative and meant not to be a direct comparison to highly-resolved GMCs. It is interesting to note that the two tidal tail regions in \PSBI with detected H$\alpha$ and thus ongoing star-forming activity have the lowest velocity dispersions of any components of either galaxy.

We examine the relationship between the velocity dispersion normalized by clump size and gas surface density in the right panel of Figure \ref{fig:turb}. 
We include a dotted line representing the expected relationship for virialized clouds, where the balance between gravitational potential energy and kinetic energy defines the gas dynamics. In this case, we expect $\Sigma_{\rm gas} \sim 370\sigma^2 / R$ where $\Sigma_{\rm gas}$ is in units of $\Msol \rm pc^{-2}$, $\sigma$ in \kms, and $R$ in pc. Structures within a factor of three (shaded band) are typically considered gravitationally bound and consistent with virial equilibrium \citep[e.g.,][]{dessauges-zavadsky_molecular_2019,rosolowsky_giant_2021}. The tail clumps in each PSB fall within this relation, although their velocity dispersions are likely dominated by non-virialized motions. Their location in this panel (and in the left) is simply used to illustrate how their kinematics differ from the centrals. The central galaxies are above the relation, exceeding what is expected for virialized structures. Instead they may be influenced by external factors such as feedback from AGN, residual merger dynamics, or strong stellar feedback, which inject energy and disrupt the gravitational binding of the molecular gas (e.g., \citealt{schinnerer_molecular_2024} and references therein).  

\subsubsection{The Role of Radio-Mode Feedback from AGN} \label{radiomode}
We can also investigate the role that AGN could play in the ongoing suppression of star formation in these PSBs. Two primary modes of AGN feedback are frequently discussed in the literature: the `quenching-mode' (or `quasar-mode') and the `maintenance-mode' (or `radio-mode'). In the quasar-mode, radiative pressure from an accreting black hole is thought to drive outflows, ejecting cold gas and preventing further star formation. In contrast, radio-mode AGN feedback involves the production of radio jets that inject mechanical energy into the interstellar medium, heating the gas and inhibiting its cooling, thereby preventing star formation \citep[e.g.,][]{croton_many_2006}. In \PSBI and \PSBII, major mergers likely triggered a starburst and subsequent quenching via the tidal removal of cold gas, but AGN-driven radio-mode feedback may help maintain the quenched state by preventing gas accretion and cooling.

The left panel of Figure \ref{fig:LradioMstar} shows the 1.4\,GHz radio luminosity against the total stellar mass for both PSBs. To enable a consistent comparison with other samples of radio AGN, we convert the 6\,GHz flux densities of the two PSBs to their corresponding 1.4\,GHz luminosities using a spectral index of $\alpha = -0.7$, assuming the typical power-law relation for synchrotron emission $S_\nu \propto \nu^\alpha$. We also show comparison samples from other AGN populations, including radio AGN in \squiggle galaxies detected in FIRST \citep{greene_role_2020}, radio AGN out to $z\sim1$ from the deep 3\,GHz VLA-COSMOS survey \citep{smolcic_cosmic_2009}, and $z\sim0.2$ PSBs hosting radio AGN \citep{shin_radio_2011}. \PSBI sits firmly within the scatter defined by these literature radio samples, while \PSBII is modestly radio weak given its large stellar mass. \citet{shin_radio_2011} suggest that the mild correlation between radio power and total stellar mass may indicate that AGN feedback is linked to the more massive, older stellar populations in galaxies. \PSBI and \PSBII fall within the scatter of this relation, suggesting that the radio emission in these two PSBs could be connected to the evolved stellar population in the galaxies.

Understanding the connection between radio AGN activity and stellar populations can provide insight into the timing of AGN feedback relative to quenching. The right panel of Figure \ref{fig:LradioMstar} shows the 1.4\,GHz radio luminosity against the fraction of stellar mass associated with the starburst $f_{\rm burst}$, motivated from \citet{shin_radio_2011} who first explored the connection between radio AGN activity and stellar populations in PSBs of this cosmic epoch in this manner. 
The two PSBs from this work are plotted alongside the $z\sim0.2$ PSBs from \citet{shin_radio_2011} where no clear correlation is observed, suggesting that the radio emission is not associated with the younger stellar populations or the starburst event that occurred $\sim$1\,Gyr ago. This implies that if the radio emission is due to radio-mode AGN, it was not active during the starburst phase and did not contribute to quenching star formation by directly suppressing the formation of new stars. Instead, the AGN may have turned on after the starburst phase during the post-starburst evolution, and is now preventing the accretion of gas and subsequent star formation. This aligns with previous findings that AGN activity often lags behind starburst-driven quenching \citep[e.g.,][]{yesuf_starburst_2014,pawlik_origins_2018}, where the AGN acts as a mechanism to maintain quiescence rather than initiate it, and consistent with the idea that AGN feedback can occur well into the post-merger stage in PSBs \citep[e.g.,][]{schweizer_o_2013}.   

Taken together, the central galaxies are not forming stars efficiently despite their substantial gas reservoirs, stemming from high levels of turbulence that stabilizes the gas against gravitational collapse. In tandem, radio-mode feedback further disrupts the star formation by injecting energy into the interstellar medium, preventing gas from cooling and forming stars and maintaining a quenched state. It is important to note that the radio emission could also originate primarily from star formation (perhaps the case for \PSBII; Figure \ref{fig:radioPredicted}), where the radio emission could reflect a period of continued, but significantly reduced, star formation following the initial burst and subsequent quenching event.  

\section{Conclusions} \label{conclusions}
In this paper we investigate the role of tidal forces in quenching star formation in two massive $z\sim0.7$ PSBs from the \squiggle sample. We present the finding of extreme molecular gas characteristics in the tidal tails of \PSBII which host $55\%\pm5\%$ of the molecular gas in the system, the second such galaxy to exhibit these features after \PSBI presented in \citetalias{spilker_star_2022}. The magnitude and spatial extent of the gas in the tidal tails significantly exceed those observed in nearby mergers, indicating that the tidal removal of gas is likely the dominant quenching mechanism in these systems. 

The discovery of vast molecular gas tidal tails in two independent PSBs suggests that this mechanism may be a common occurrence in intermediate-redshift PSBs. Out of the six CO-detected PSBs from \citet{bezanson_now_2022}, two exhibit these extended gas features, indicating that such structures could be a relevant component of the quenching process in this population. Molecular mass scales with redshift as $(1+z)^a$ where $a \approx 2.5-3$ \citep[e.g.,][]{tacconi_evolution_2020} and the rate of galaxy mergers increases as $(1+z)^m$ where $m \approx 2-3$ \citep[e.g.,][]{lotz_evolution_2008,conselice_structures_2009,bridge_cfhtls-deep_2010,duncan_observational_2019, ferreira_galaxy_2020}. Thus, the frequency of mergers and availability of gas in high-redshift systems could explain the presence of such gas-rich tidal structures. 

We show that the star-forming properties of the tidal tails differ between the two galaxies. While \PSBI exhibits H$\alpha$ emission in its northern tidal tail consistent with ongoing star formation corresponding to $\sim$5\,\sfr, the tidal tails of \PSBII lack detectable H$\alpha$, underscoring the complexity of gas dynamics and star formation in tidally removed gas. Tidal debris in nearby massive galaxies often shows signs of star formation that later coalesces into structures such as TDGs \citep[e.g.,][]{duc_young_1998,braine_formation_2000,duc_formation_2000}, suggesting that the star-forming clumps in the tidal regions of \PSBI could experience a similar fate long-term. 

Although they possess substantial molecular gas reservoirs the central galaxies of both PSBs are inefficient at forming stars, with depletion times an order of magnitude larger than typical galaxies at $z\sim1$. This inefficiency is attributed to high turbulence ($\sigma > 200$\,\kms) in the molecular gas, stabilizing it against gravitational collapse. Additionally, evidence of AGN activity particularly in \PSBI suggests that radio-mode feedback may play a maintenance role by heating the gas in the ISM and preventing its cooling, further inhibiting star formation \citep[e.g.,][]{croton_many_2006}. This scenario aligns with previous findings which suggest that AGN activity often lags behind quenching and acts as a mechanism to maintain the suppression of star formation rather than act as the driving force \citep[e.g.,][]{yesuf_starburst_2014,pawlik_origins_2018}.

Despite broad similarities, \PSBI and \PSBII exhibit differences in the exact scale, alignment, and star-forming properties of their tail features. These differences likely reflect variations in merger dynamics and gas dissipation, as well as the influence of ancillary mechanisms such as AGN feedback or environmental effects. However with only two objects studied thus far, we have yet to sample the full parameter space of tidal gas removal. Expanding the sample to include a statistically significant number of galaxies across a range of cosmic epochs will be important in understanding these variations. Future mid-J CO observations and improved constraints on dust continuum emission in the tails will help refine the physical conditions of the molecular gas, while deeper H$\alpha$ and Pa$\alpha$ observations will better constrain residual star formation and ultimately the fate of the tidally removed gas.

\begin{acknowledgements}

V.R.D. acknowledges Gabe Brammer and Jasleen Matharu for their valuable assistance with the HST grism data reduction. 
V.R.D., J.S.S., and K.A.S. gratefully acknowledge support from the National Science Foundation under NSF-AAG No. 2407954 \& 2407955. 
V.R.D. also acknowledges support provided by the NSF through award SOSPA 11-006 from the NRAO.
This paper makes use of the following ALMA data: ADS/JAO.ALMA\#2017.1.01109.S, ADS/JAO.ALMA\#2018.1.01264.S, ADS/JAO.ALMA \#2019.1.00221.S. ALMA is a partnership of ESO (representing its member states), NSF (USA) and NINS (Japan), together with NRC (Canada), MOST and ASIAA (Taiwan), and KASI (Republic of Korea), in cooperation with the Republic of Chile. The Joint ALMA Observatory is operated by ESO, AUI/NRAO and NAOJ. The National Radio Astronomy Observatory is a facility of the National Science Foundation operated under cooperative agreement by Associated Universities, Inc. 
This research is based on observations made with the NASA/ESA \textit{Hubble Space Telescope} obtained from the Space Telescope Science Institute, which is operated by the Association of Universities for Research in Astronomy, Inc., under NASA contract NAS 5–26555. These observations are associated with programs 16201 and 16248. All of the data presented in this paper were obtained from the Mikulski Archive for Space Telescopes (MAST) at the Space Telescope Science Institute. The specific observations analyzed can be accessed via \dataset[https://doi.org/10.17909/9xpd-3752]{https://doi.org/10.17909/9xpd-3752}. Support to MAST for these data is provided by the NASA Office of Space Science via grant NAG5–7584 and by other grants and contracts. 
This research has made use of NASA's Astrophysics Data System.

\end{acknowledgements}

\facility{ALMA, HST (WFC3), Sloan, VLA}

\software{
\texttt{astropy} \citep{astropy_collaboration_astropy_2018},
CASA \citep{mcmullin_casa_2007},
\texttt{Grizli} \citep{brammer_grizli_2019},
\texttt{matplotlib} \citep{hunter_matplotlib_2007}}

\appendix

\section{Residual Analysis of \PSBII} \label{J2258Residuals}

\begin{figure*}[ht!]
\centering{ 
\includegraphics[width=0.85\textwidth]{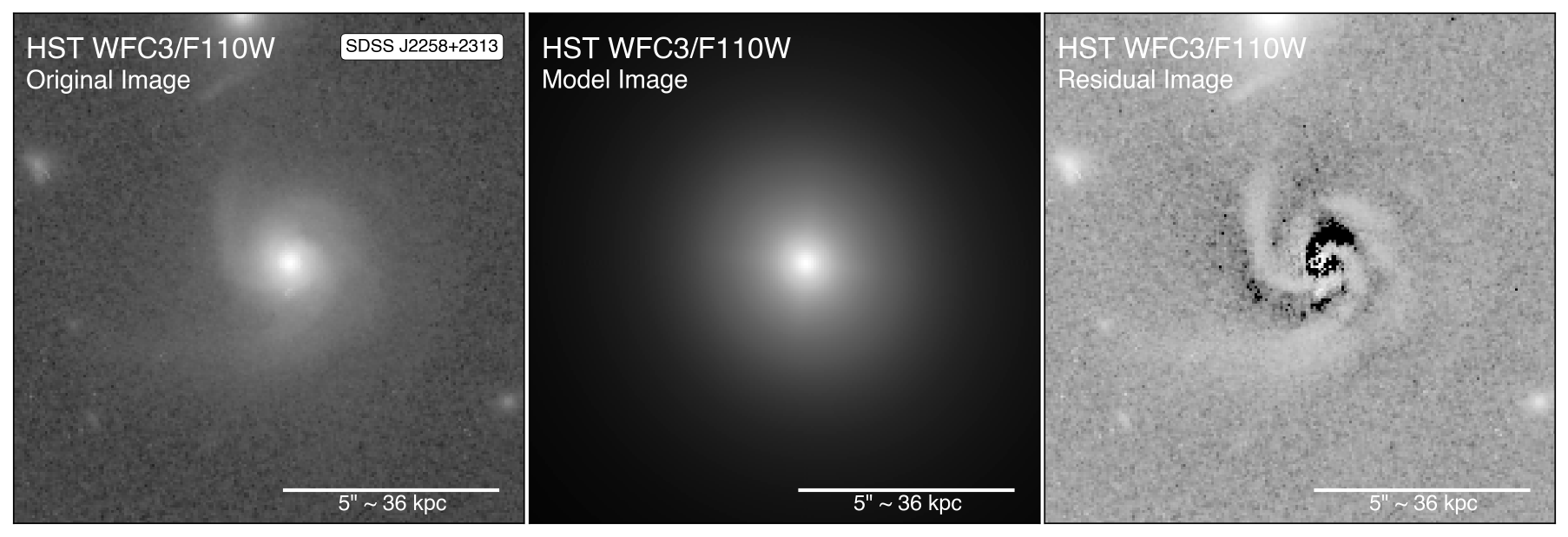}
}
\caption{HST WFC3/F110W imaging and S\'{e}rsic model fitting results for \PSBII, with the original HST image (left), best-fit single S\'{e}rsic model (middle), and residual image (right). The asymmetries in the tidal features and the minor peak of light located $\sim$0.5\arc southward from the galaxy center is emphasized by the residuals.
}\label{fig:J2258Residuals}
\end{figure*}

Figure \ref{fig:J2258Residuals} presents the results of the structural modeling of \PSBII using HST WFC3/F110W imaging. The left panel displays the original F110W image. To characterize the galaxy's structure, we fit a two-dimensional S\'{e}rsic profile to the galaxy using the \texttt{pysersic} package \citep{pasha_pysersic_2023}. During the fitting process we masked nearby sources to prevent contamination in the model fitting. The best-fit model is shown in the middle panel. The right panel presents the residual image, obtained by subtracting the best-fit S\'{e}rsic model from the original F110W image. A notable feature is the minor peak in the light distribution near the center of the galaxy, at the base of the northern tail. This peak is offset from the galaxy center by $\sim$0.5\arc, or $\sim$3.6\,kpc at the redshift of \PSBII, suggesting a possible merging companion that has not yet fully coalesced with the host galaxy. Additionally, residuals trace the location of the asymmetric extended tidal features, further reinforcing the interpretation that the disturbed morphology and elongated structures are the result of recent tidal interaction. 

\section{CO(4--3) Observations of \PSBI} \label{CO43}

\begin{figure*}[h!]
\centering{ 
\includegraphics[width=0.85\textwidth]{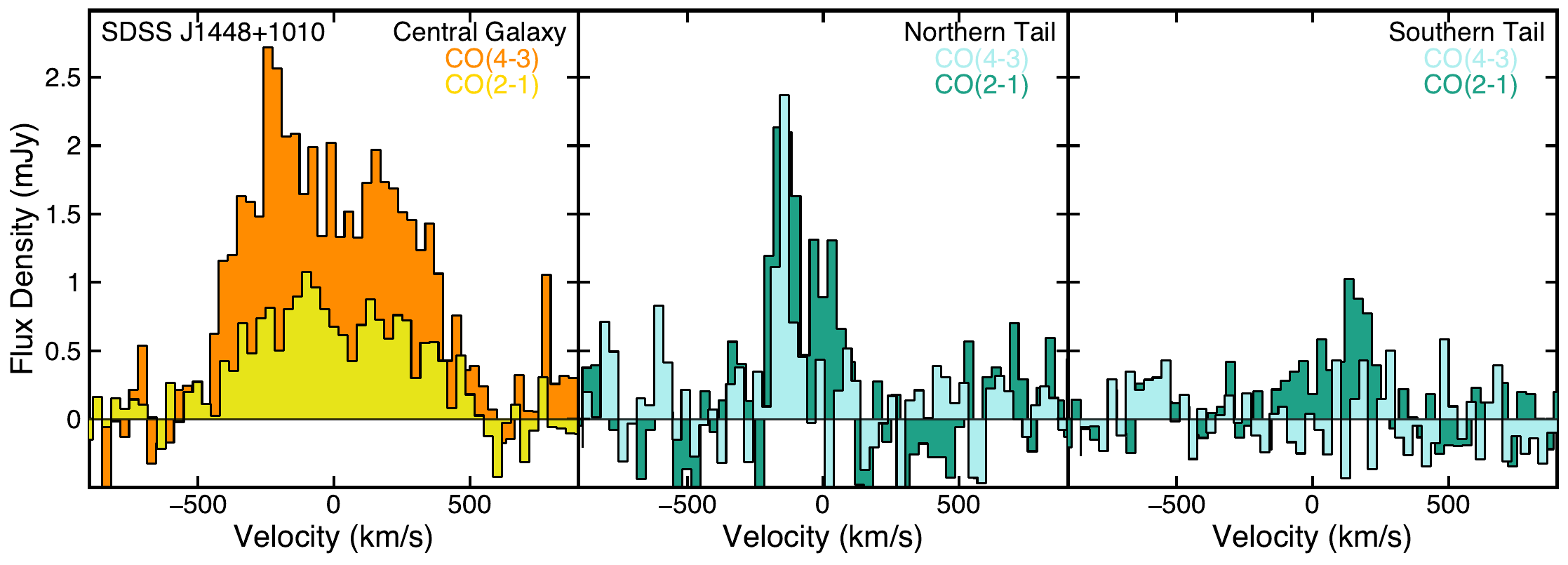}
}
\caption{ALMA CO(4--3) and CO(2--1) spectra for the central region (left), northern tail (middle), and southern tail (right) of \PSBI. The CO(4--3) emission is detected in the central galaxy and northern tail with greater intensity compared to CO(2--1), but is not detected in the southern tail.
}\label{fig:COSpectraJ1448with43}
\end{figure*}

In addition to the CO(2--1) data presented in this work, we also obtained observations of the CO(4--3) transition for \PSBI, targeting the same regions as those covered by the CO(2--1) observations: the central region and northern and southern tidal tails (PI: Suess, Program \#2018.1.01264.S). The CO(4--3) observations were carried out using a similar imaging strategy described in Section~\ref{almaobs}, but we do not include this dataset in the main analysis due to the lack of comparable CO(4--3) observations for \PSBII.

Figure \ref{fig:COSpectraJ1448with43} shows the CO(4--3) spectra for the central galaxy and the northern and southern tidal tails in \PSBI. Similar to the CO(2--1) data presented in Figure \ref{fig:momentsSpecSDSSJ1448}, we detected CO(4--3) emission in the central region and northern tail, with the central galaxy exhibiting a broader profile compared to the narrower profiles observed in the tails. However, CO(4--3) emission is not detected in the southern tail feature. The ratios in flux units for the CO(4--3) and CO(2--1) lines are r$_{42}=2.43\pm0.13$ for the central component, r$_{42}=0.32\pm0.13$ for the northern component, and r$_{42}<0.87$ (3$\sigma$ upper limit) for the southern component. 

\bibliographystyle{aasjournal}
\bibliography{QuenchingviaTidalRemoval.bib}

\end{CJK*}
\end{document}